\documentclass[acmsmall]{acmart}
\AtBeginDocument{%
  }
\renewcommand\footnotetextcopyrightpermission[1]{}
\usepackage{multirow}
\setcopyright{acmlicensed}
\copyrightyear{2018}
\acmYear{2026}
\acmDOI{XXXXXXX.XXXXXXX}

\acmJournal{TACO}
\acmVolume{0}
\acmNumber{0}
\acmArticle{0}
\acmMonth{0}

\usepackage[english]{babel}
\usepackage{caption}
\usepackage{subcaption}

\usepackage[dvipsnames]{xcolor}
\usepackage{amsmath}
\usepackage{graphicx}
\usepackage{multirow}

\usepackage{algorithm}
\usepackage{algpseudocode}

\usepackage{booktabs}   
\usepackage{tabularx}   

\begin{document}


\author{Yuxuan Wang}
\authornote{Yuxuan Wang and María José Belda are co-first authors.}
\email{yuxuan.wang@epfl.ch}
\affiliation{%
  \institution{EPFL}
  \city{Lausanne}
  \country{Switzerland}
}

\author{Mar\'{i}a Jos\'{e} Belda}
\authornotemark[1]
\email{mbelda@ucm.es}
\affiliation{%
  \institution{Universidad Complutense de Madrid}
  \city{Madrid}
  \country{Spain}
}

\author{Fernando Castro}
\email{fcastror@ucm.es}
\affiliation{%
  \institution{Universidad Complutense de Madrid}
  \city{Madrid}
  \country{Spain}
}

\author{Katzalin Olcoz}
\email{katzalin@ucm.es}
\affiliation{%
  \institution{Universidad Complutense de Madrid}
  \city{Madrid}
  \country{Spain}
}

\author{David Atienza}
\email{david.atienza@epfl.ch}
\affiliation{%
  \institution{EPFL}
  \city{Lausanne}
  \country{Switzerland}
}

\author{Giovanni Ansaloni}
\email{giovanni.ansaloni@epfl.ch}
\affiliation{%
  \institution{EPFL}
  \city{Lausanne}
  \country{Switzerland}
}

\renewcommand{\shortauthors}{Wang et al.}

\def\titleacronym{kernel}
\def\mmul{\texttt{mmul}}
\def\Mmul{\texttt{Mmul}}
\def\mmuls{\texttt{mmuls}}

\title{Exposing Embedded Matrix Multiplication with polyhedral transformations for CGRA compilation}

\renewcommand{\shortauthors}{Wang et al.}

\begin{abstract}

Matrix-matrix multiplication (\mmul{}) is a dominant computing pattern in edge workloads, but in real applications it is often embedded within larger affine programs and intertwined with other operations. Coarse-Grained Reconfigurable Arrays (CGRAs) can flexibly and efficiently support it, since they combine operation-level reconfigurability and high energy efficiency. However, mapping computational kernels that include \mmul{} with state-of-the-art compilation strategies often leads to suboptimal results, since its multi-dimensional structure hampers the uncovering of its inherent parallelism and, ultimately, runtime performance. This paper presents Kernel-CGRA, a compiler flow that uses polyhedral transformations to expose legal affine \mmul{} subspaces, lower each accepted region to a back-end-independent operation, and preserve the surrounding computation in the generic CDFG path. For CGRA execution, the operation is instantiated as a parametric output-stationary kernel, while residual regions are reconnected through context generation and live-value preservation. The matcher is intentionally strict: only regions proven to match the supported affine \mmul{} form are rewritten. Across canonical, fused, and application-level benchmarks, Kernel-CGRA achieves runtime speedups of up to $10.2\times$ over state-of-the-art CGRA CDFG compilation with modulo scheduling and loop unrolling.
\end{abstract}



\received{5 June 2026}
\received[revised]{1 September 2026}
\received[accepted]{5 October 2026}

\maketitle

\section{Introduction}
Matrix multiplication is a core computation in edge workloads, including machine learning (ML) operations such as attention, fully connected layers, and convolutions, as well as data processing and signal analysis~\cite{karstadt2020matrix}. At the same time, real edge applications are not limited to matrix multiplication alone; they often combine it with additional computations in complex pipelines~\cite{suhre2013multiplication}. As ML is increasingly deployed on edge devices for healthcare, autonomous driving, and industrial automation~\cite{murshed2021machine,marculescu2020edge}, these workloads must be executed efficiently under tight resource constraints~\cite{ace}. However, the constrained energy and computation budgets of edge platforms pose significant challenges in the execution of computation-intensive tasks~\cite{7879258}. In this context, CGRAs offer a compelling solution. They are programmed at the operation level, offering a trade-off between performance and flexibility, while maintaining adaptability compared to Application-Specific Integrated Circuits (ASICs) and delivering higher energy efficiency than general-purpose processors~\cite{li2022coarse,akbari2019x,prasad2025ultra}.

A CGRA typically consists of multiple processing elements (PEs) interconnected in a grid-like structure~\cite{cgra-me}. The PE arrays enable parallel execution of multiple operations, with data values propagated between connected PEs to satisfy data dependencies. Consequently, the key challenge in CGRA deployment is compilation: routing resources must be used efficiently to resolve dependencies while maximizing instruction-level parallelism~\cite{20years_cgra}.

\begin{figure}
    \centering
    \includegraphics[width=0.6\textwidth]{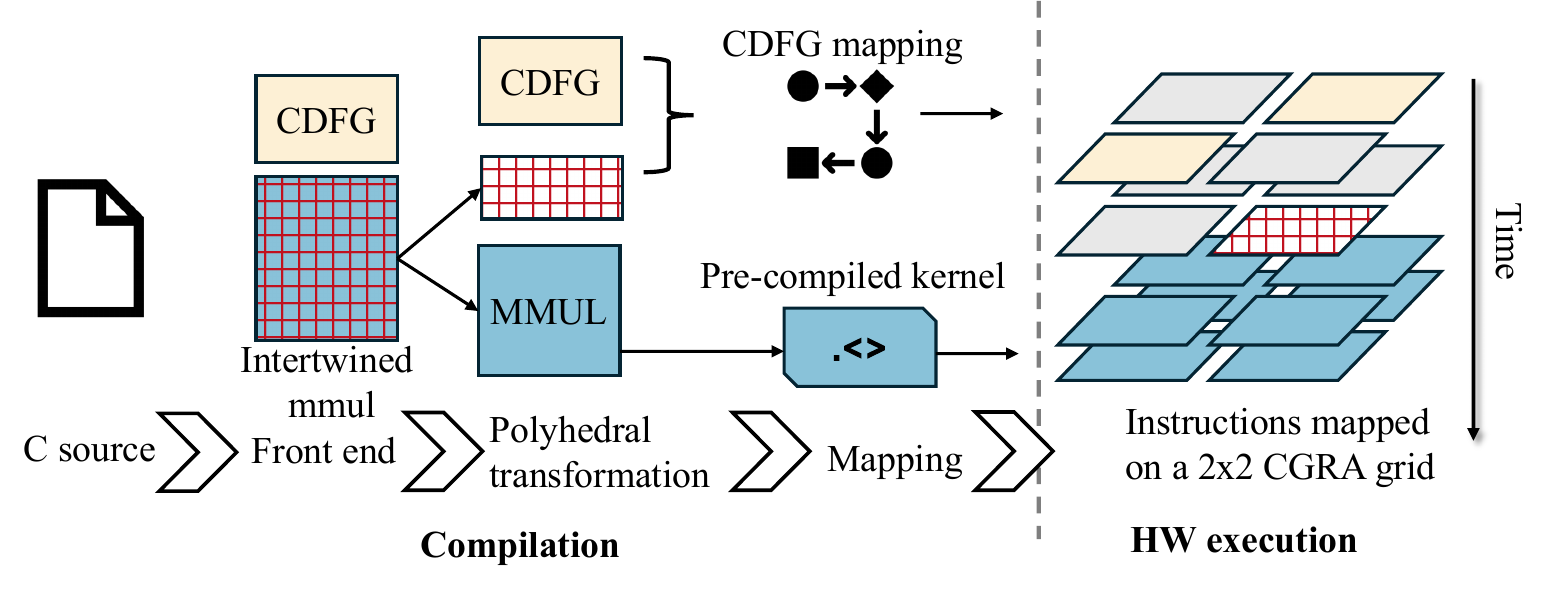}
    \caption{Overview of the proposed CGRA compilation flow integrating pre-compiled kernels for efficient hardware execution.}
    \vspace{-0.5em}
    \label{fig:cgra hw}
    \vspace{-1em}
\end{figure}

This compilation problem is addressed by mapping operations in space and scheduling them in time, usually from static single-assignment representations of user code generated by compiler infrastructures such as LLVM/MLIR~\cite{MLIR}.
Modulo scheduling (MS) is the primary compilation approach, mapping a control-free loop's dataflow graph (DFG) onto hardware. Its goal is to minimize the initiation interval (II), i.e., the time gap between successive loop iterations, to maximize the execution of parallel operations during steady-state computation~\cite{cgra-me,satmapit_date, Mei2007}. Recent compilation techniques have expanded the scope beyond control-free loops to support control-dataflow graphs (CDFGs)~\cite{wang2025cfg, 10.1145/3808285}. While broadening the scope of compilable programs, they still rely on MS to optimize computation-intensive loop kernels, thereby improving runtime performance.

Although there are many automated compilers, a significant performance gap remains between automated mapping and manual mapping, both in terms of quality and scope~\cite{cgra-me,morpher}. Indeed, prior work has demonstrated that manual mapping for specialized ML kernels can enable efficient execution on CGRAs~\cite{ml_cgra}. However, such approaches are typically tailored for specific operations, such as convolutions, and cannot be generalized to broader data-processing applications on edge devices.
Automated approaches, such as MS, can target diverse applications, but their performance is constrained by their limited scope and is not well-suited for matrix multiplication, which can be parallelized at different loop levels. 

To fully leverage this structure while preserving CGRA flexibility, we propose Kernel-CGRA, a framework that changes the compilation boundary for \mmul-dominated programs. Instead of mapping only operations visible in a basic block or innermost loop, Kernel-CGRA uses polyhedral transformations to expose legal affine \mmul{} subspaces inside larger programs and rewrites each accepted subspace as an explicit \texttt{cgra.mmul} operation. As shown in Figure~\ref{fig:cgra hw}, exposed \mmul{} regions use optimized pre-compiled kernels, while the remaining computation (yellow/dashed regions) stays in the regular CDFG path. This split preserves generality while enabling data reuse and outer-loop parallelism. The extracted \mmul{} representation is back-end independent: in our evaluation, both the proposed CGRA kernel and the SA baseline consume the same exposed \mmul{} regions, so the comparison isolates back-end execution and residual-code integration.

Our strategy resembles library-based approaches such as MKL and cuBLAS~\cite{intel_mkl,nvidia_cublas} for typical CPUs and GPUs, but those approaches typically rely on explicit calls or source-level kernel boundaries. Kernel-CGRA instead covers \mmul{} subspaces through compiler analysis even when they are embedded in larger affine computations, while keeping the surrounding code context.

In summary, the main contributions of the paper are as follows:

\begin{itemize}
      \item A parametric output-stationary \mmul{} kernel for homogeneous CGRAs, designed to exploit spatial data reuse across rows and columns while remaining configurable across CGRA sizes.
      \item A polyhedral middle-end that exposes intertwined affine \mmul{} subspaces through scalarization, reduction distribution, memory-reference canonicalization, and statements reordering.
      \item A backend-independent \texttt{cgra.mmul} representation that separates kernel exposure from target-specific lowering, enabling comparison against different accelerator backends.
      \item A kernel-integration mechanism that preserves program semantics across extracted regions through context generation and live-value spill/restore, allowing residual non-\mmul{} code to remain in the CDFG compilation path.
      \item An end-to-end evaluation showing runtime speedups of up to $10.2\times$ over state-of-the-art CGRA CDFG compilation with MS and loop unrolling.
\end{itemize}

The paper is organized as follows. Section~\ref{sec:related} reviews related work and Section~\ref{sec:background} introduces the CGRA execution model and motivating limitations of existing compilation flows. Section~\ref{sec:method} presents the overall Kernel-CGRA methodology, while Section~\ref{sec:compilation} details the middle-end transformations used to expose \mmul{} subspaces. Section~\ref{sec:backend} describes kernel generation and integration with residual code. Section~\ref{sec:exp} evaluates the proposed flow and reports runtime speedups and scalability. Section~\ref{sec:ablation} conducts an ablation study to isolate the impact of individual middle-end passes and backend kernel substitution. Finally, Section~\ref{sec:conclusion} concludes the paper.

\section{Related work}
\label{sec:related}

\subsection{Compilation for CGRA}
\label{subsec:cgra-compilation}

The key challenge in compiling to CGRAs lies in mapping applications onto hardware resources, scheduling operations in space (PE assignment) and time (clock cycle). Most CGRA compilers focus on efficiently mapping loops~\cite{20years_cgra, wang2025cfg}. In this context, MS is the most common approach that exploits the parallelism of CGRA by pipelining and overlapping successive loop iterations, accounting for loop-carried data dependencies~\cite{robust_compilation,concrrent_modulo,MonoMap}. Various MS implementations have been explored~\cite{satmapit,riptide}, based on optimization approaches such as satisfiability formulations and Integer Linear Programming (ILP). 
Nevertheless, MS strategies are ill-suited for the \mmul{} kernel: although  MS compilers can apply scheduling to the innermost loop, they fail to fully exploit the  parallelism opportunities beyond it, i.e., in \mmul{}  outer loops. Mitigating this shortcoming, ADRES~\cite{Mei2007} supports loop unrolling to expose additional parallelism in inner loops. However, such a strategy only offers a partial solution, as in general \mmul{}s cannot be completely flattened without a large increase in the number of operations to be scheduled, which itself is a source of inefficiencies ~\cite{wang2025cfg}.

\begingroup
\setlength{\textfloatsep}{0.4em}
\setlength{\belowcaptionskip}{-0.5em}
\begin{figure}
    \centering
    \includegraphics[width=0.8\textwidth]{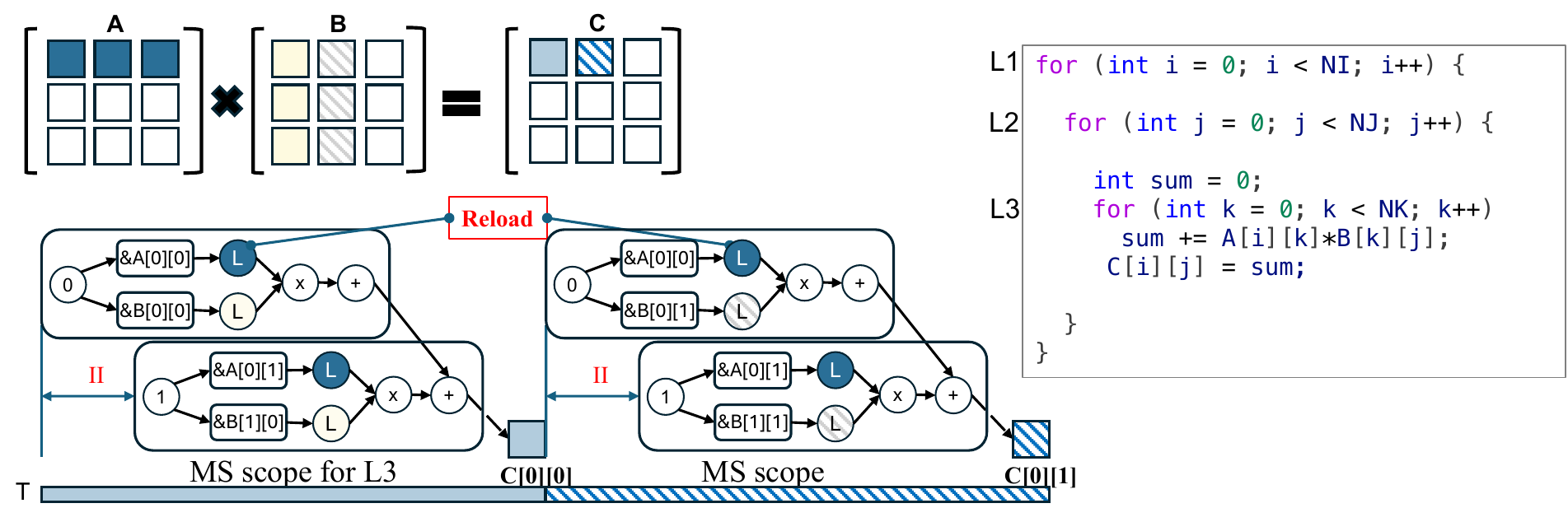}
    \caption{Execution flow of \mmul{} based on CDFG compilation, highlighting the inefficiency resulting from only exploiting data parallelism in innermost loops.}
    \label{fig:how ms}
\end{figure}
\endgroup

Beyond the scheduling of single loops, CDFG mappings extend the compilation scope to arbitrary control-flow graphs. Such schedulers \cite{wang2025mlir} can target complex kernels, limiting communication overhead and its associated latency between the CGRA and host processors. Moreover, they can embed MS as a component to derive highly efficient schedules for inner loops. However, similarly to MS-based compilers, CDFG-based approaches also map operations on each basic block  (e.g. that of a loop body) independently,  which prevents them from breaking loop hierarchies. As a result, higher-level loops cannot begin execution until the inner loops complete, limiting opportunities for cross-level parallelism and preventing effective exploitation of the abundant data reuse in \mmul.

Highlighting the resulting inefficiency, Figure~\ref{fig:how ms} illustrates the execution process generated by CDFG-based methods~\cite{wang2025mlir} that apply MS to optimize the runtime of the innermost loop $L3$. By minimizing the II, these methods allow the partial products of \mmul{} to be computed in parallel, as shown in the MS scope for $L3$. However, although the computations of different output elements are independent, MS cannot exploit this outer-loop parallelism. In the generated CDFG, each outer-loop iteration must wait for its corresponding inner loop to complete before proceeding ($j=1$ after $j=0$). Note that loop unrolling can alleviate, but not completely solve, this issue. Furthermore, address generation and data loading for input matrices impose significant overhead, as control-flow constraints prevent the reuse of loaded data across multiple output computations.

To exploit the parallelism of the \mmul{} kernel and boost performance, a common approach adopted across different architectures is to use specialized, precompiled libraries tailored to specific hardware. For instance, Intel’s oneMKL offers tuned CPU variants of Basic Linear Algebra Subprograms (BLAS)~\cite{intel_mkl}, and cuBLAS provides highly optimized BLAS operations on NVIDIA GPUs~\cite{nvidia_cublas}. These libraries demonstrate the effectiveness of pre-compiling domain-specific kernels to exploit hardware capabilities and achieve high performance for fixed computation patterns.

We adopt a similar strategy, but targeting CGRAs. Several CGRA frameworks also rely on specialized kernels or accelerator templates: ML-CGRA uses pre-compiled ML operators~\cite{ml_cgra}; CGRA4ML targets neural-network deployment for scientific edge computing~\cite{cgra4ml_trets}; CFEACT supports CNN and transformer accelerator design~\cite{cfeact}; and systolic mapping on elastic CGRAs focuses on efficient matrix multiplication and reduction~\cite{systolic_elastic_cgra}. These methods benefit their target domains through optimized dataflow, hardware/software integration, but their scope is mainly ML-oriented accelerator generation or standalone dense-kernel execution.
In contrast, our method targets a broader compilation problem: it uses polyhedral analysis to expose legal \mmul{} subspaces, maps those regions to parametric pre-compiled kernels, and integrates them with CDFG compilation for the remaining program. As a result, it benefits from kernel-level data reuse and latency-aware scheduling while preserving automated CGRA compilation for applications where \mmul{} is embedded in larger affine and control-flow contexts.

\vspace{-0.5em}
\subsection{Polyhedral compilation}
Polyhedral representations model loop iterations and data accesses using affine relations. This representation enables structured optimizations to improve data locality and parallelism. Polyhedral transformations are widely used in the context of multi-core and SIMD architectures~\cite{polyhedral_simd}. More recently, they have been adopted for operation scheduling in high-level synthesis~\cite{polyhedral_hls} and in CGRA compilation to handle imperfect loop executions~\cite{polyhedral_cgra}. 
However, these approaches primarily focus on loop scheduling and transformations and therefore lack domain-specific knowledge of data-sharing and reuse patterns when compiling for specific \mmul{} kernels.
Conversely, we use polyhedral transformations to expose \mmul{} kernel patterns - nested-loop structures forming a core component of many computation-intensive tasks - which are integrated into the compilation of entire CDFGs, which may contain \mmul{} and non-\mmul{} regions.

\section{Background and motivation}
\label{sec:background}

\subsection{Polyhedral model}
The polyhedral model represents affine loop nests as integer points and relations~\cite{polyhedral_basics}. Iteration domains enumerate dynamic statement instances, access functions map those instances to memory locations, dependence relations constrain legal reordering, and schedule functions define execution order.
To make these ideas concrete, we summarize these four core concepts below.

\vspace{-0.5em}
\subsubsection{Iteration domain and access functions}
The iteration domain represents the set of all dynamic executions of a statement. Each time the statement is executed corresponds to a specific point in this domain. In Figure~\ref{fig:background}, each point $(i,j)$ in the 2D grid represents one execution of statement $S_1$, and the arrows indicate the iteration order. The iteration domain is bounded on $i$ and $j$:

\begingroup
\vspace{-1em}
\setlength{\abovedisplayskip}{-2pt}
\setlength{\belowdisplayskip}{-2pt}
\begin{align*}
\mathcal{D}_{S_1} = \{(i,j)\in\mathbb{Z}^2 \mid 0 \le i < N_I,\; 0 \le j < N_J\}
\end{align*}
\endgroup

An access function maps an iteration point to the memory element touched by an array reference, i.e., it formalizes which element each loop instance reads or writes. In Figure~\ref{fig:background} example, for the statement $S_1$ the read $C[i][j]$ is
\begingroup
\setlength{\abovedisplayskip}{-2pt}
\setlength{\belowdisplayskip}{-2pt}
\begin{align*}
F^{S_1}_C
&=
\begin{pmatrix}
1 & 0 \\
0 & 1
\end{pmatrix}
\begin{pmatrix}
i & j 
\end{pmatrix}^{T}
=
\begin{pmatrix}
i & j
\end{pmatrix}^{T}.
\end{align*}
\endgroup
Similarly, $v$ has two reads in the same $S_1$ instance, expressed as $F^{S_1}_{v,1}(i,j)=i$ and $F^{S_1}_{v,2}(i,j)=j$. 

\vspace{-0.5em}
\subsubsection{Data dependencies}
Data dependencies capture ordering constraints between statement instances that access the same memory location. For a dependence $S_p\rightsquigarrow S_q$, constrained pairs are represented by $\mathcal{R}_{p\rightarrow q}\subseteq\mathcal{D}_{S_p}\times\mathcal{D}_{S_q}$, and each $(d_p,d_q)\in\mathcal{R}_{p\rightarrow q}$ requires $d_q$ to execute before $d_p$.

\vspace{-0.8em}
\subsubsection{Schedule functions}
A schedule assigns to each statement instance a multidimensional timestamp. From a schedule, an execution order is derived following the lexicographic order of these timestamps. 
For example, the ordering constraint $\Theta_{S_1}(i,j,k,1)^T \prec \Theta_{S_0}(i,j,1)^T$ for initialization $S_0$ and accumulation $S_1$ places all accumulation instances before finalization. The scheduler searches for a legal order that preserves dependencies and reshapes the loop structure to match the target kernels. Given dependence relation $S_p\rightsquigarrow S_q$, legality requires:

\begingroup
\setlength{\abovedisplayskip}{4pt}
\setlength{\belowdisplayskip}{2pt}
\begin{align*}
\Theta^{S_p}d_p \preceq \Theta^{S_q}d_q,
\quad \forall (d_p,d_q) \in \mathcal{R}_{p\rightarrow q},
\end{align*}
\endgroup
where $\mathcal{R}_{p\rightarrow q}$ is the dependence polyhedron. This compact constraint captures RAW, Write-After-Read (WAR), and Write-After-Write (WAW) ordering for legal reordering and loop splitting.

\subsection{Motivation}

\begin{figure*}[ht!]
    \centering
    \includegraphics[width=1\linewidth]{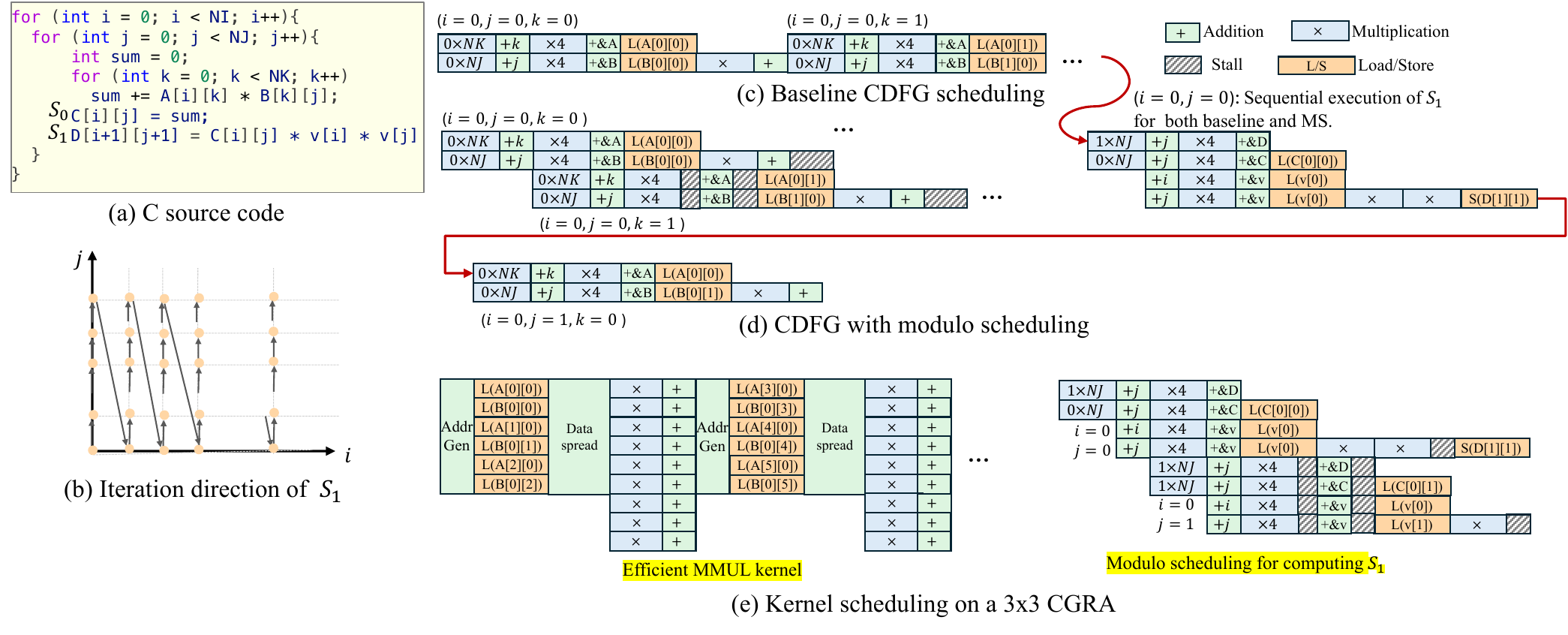}
    \caption{Motivating example for mapping a benchmark with hidden \mmul{} onto CGRA. The \textbf{baseline } exploits operations parallelism only for instructions in the same basic block. In \textbf{CDFG with MS} loop-level parallelism is used to pipeline iterations. The \textbf{proposed kernel scheduling} highly parallelise schedule, achieved by substituting the \mmul{} subspace with a pre-compiled kernel. 
    }
    \label{fig:background}
    \vspace{-1em}
\end{figure*}

Following the polyhedral formulation above, we use dependence-preserving loop transformations to expose a kernel subspace. Figure~\ref{fig:background} shows three timing views of the same program scheduled by different strategies: (1) baseline CDFG scheduling that enables instruction-level parallelism (Figure~\ref{fig:background}.c), (2) CDFG with MS with loop-level parallelism for innermost loops (Figure~\ref{fig:background}.d), and (3) our proposed kernel scheduling (Figure~\ref{fig:background}.e).

In baseline CDFG scheduling, each statement is lowered into fine-grained address generation, memory, and arithmetic operations. The arithmetic computation takes only a small portion of the total execution time, while data access with address computation dominates. CDFG MS hides part of the data-access time by folding the innermost loop. However, the compiler assumes static operation timing and is unaware of runtime hardware conditions. When operations experience different effective latencies (e.g., memory delays), the CGRA execution must stall to synchronize operations that execute in parallel, as shown in the grey-slash block in Figure~\ref{fig:background}.d.

The efficiency of our dedicated and parametric \mmul{} mapping is improved with respect to CDFG scheduling in several aspects:
1. \textbf{Latency-aligned scheduling:} Operations with the same latency are grouped into the same instruction cycle to avoid pipeline stalls. 
2. \textbf{Hybrid address generation.} Address computation is split between compile time and runtime. Affine offsets and access patterns are precomputed during compilation, while runtime only performs lightweight base-address selection, eliminating repeated index arithmetic.
3. \textbf{Full CGRA parallelism:} The kernel exposes the inherent parallelism of \mmul. As shown in Figure~\ref{fig:background}, nine Multiply-ACcumulate (MAC) operations execute concurrently on a $3\times3$ CGRA.
4. \textbf{Reuse of data from cross-iteration:} Data are reused aggressively across computations. For instance, $A[0][0]$ contributes to $C[0][0]$, $C[0][1]$, and $C[0][2]$, so it is loaded once and broadcast in hardware.
5. \textbf{Loop-structure optimization:} Polyhedral transformations restructure the execution hierarchy. By extracting the \mmul{} region as a kernel and applying loop splitting, the kernel loop becomes the innermost loop, enabling effective pipelining and loop folding across iterations.

Note that in CDFG scheduling without polyhedral transformations (Figure~\ref{fig:background}.(c) and (d)), the execution of $S_1$ cannot be parallelized, as only the innermost control-free loop can be optimized via MS, while the outer loops remain sequential. However, after extracting \mmul{} from a pre-compiled kernel (Fig. 3(e)), the independent $S_1$ become the innermost loop and can exploit loop parallelism.

\section{Methodology Overview}
\label{sec:method}

\begin{figure*}[htb!]
    \centering
    \includegraphics[width=1\textwidth]{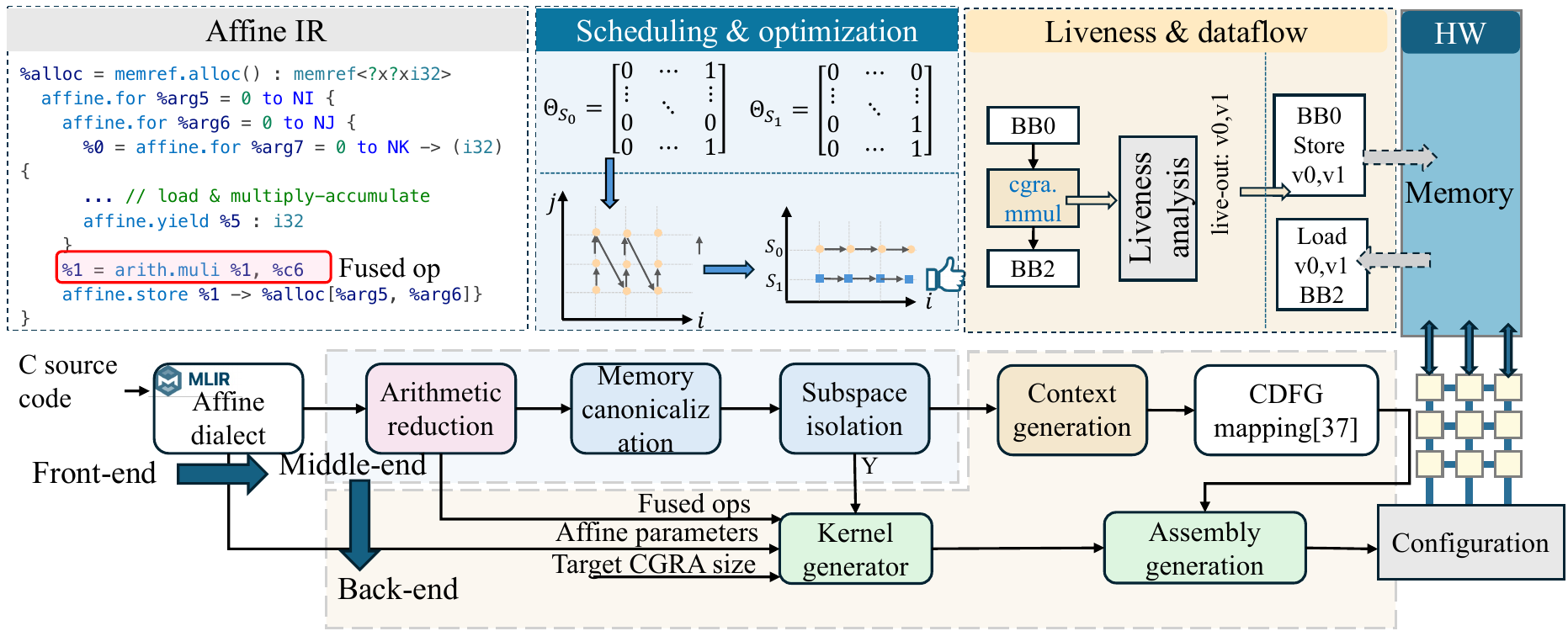}
    \caption{Overview of the integrated CGRA compilation framework: front-end (C parsing and MLIR/affine lowering), middle-end (transformations for \mmul extraction), and back-end (assembly integration with kernel instantiation).}
    \vspace{-1.5em}
    \label{fig: method overview}
\end{figure*}

A bird-eye view of our framework is shown in Figure~\ref{fig: method overview}. As depicted in the figure, it is divided into front-end, middle-end, and back-end stages, enabling the compilation of kernels from C source code to their deployment as binary configurations to CGRA hardware.

In the front-end, the input C program is parsed and lowered through MLIR to the affine dialect~\cite{MLIR}. This representation explicitly captures loop bounds, affine memory accesses, and data dependencies at the intermediate representation (IR) level, enabling subsequent transformations.

In the middle-end, linearity-based fusion, statement reordering/loop splitting, and kernel extraction with context generation are applied. First, operation fusion hoists loop-invariant terms and removes redundant memory accesses. Second, statement reordering and loop splitting solve an ILP model to isolate a legal \mmul{} computation subspace while preserving dependencies; this step is recursively applied until no further valid split is found. Third, kernel extraction matches the transformed affine form to the \mmul{} pattern. If the kernel is found, the kernel generator utilizes the collected kernel parameters (iteration domains, access parameters, and target CGRA size) to generate optimized assembly. Non-\mmul{} regions are reduced to control-flow IR on static single assignment (SSA) blocks, and liveness analysis spills and restores values that must survive kernel execution to maintain the original program semantics. 

The back-end combines the two compilation paths: the extracted \mmul{} regions are emitted as optimized kernel assembly, and the remaining non-\mmul{} control-flow regions are compiled with CDFG mapping. 

The middle-end transformations and the resulting cgra.mmul representation are independent of the concrete CGRA implementation. Hardware-specific information is introduced during back-end lowering, where each cgra.mmul operation is instantiated according to the target architecture and ISA.
The emitted assembly is integrated into a unified CGRA configuration and executed on the time-multiplexed CGRA fabric. At the end of each basic block, a branch determines the successor and transfers control to the corresponding instruction region, enabling residual basic blocks and MMUL kernels to execute sequentially following the original program control flow.

\vspace{-0.5em}
\section{Middle-end Transformations}
\label{sec:compilation}
The proposed middle-end exposes affine subspaces that are semantically equivalent to \mmul{} and lowers them to the parametric \texttt{cgra.mmul} operation, while preserving the remaining program for CDFG compilation. 
Table~\ref{tab:compilation_transformations} summarizes the implemented middle-end passes for transforming the IR and exposing and integrating \mmul{} subspaces; starred entries are provided by the MLIR infrastructure. The middle-end first normalizes affine reductions, then exposes and verifies legal \mmul{} subspaces, and finally lowers the residual program while preserving the extracted kernel interface. The following subsections describe these transformations in detail.

\begin{table}[t]
\caption{Implemented middle-end passes for exposing and integrating \mmul{} subspaces.}
\label{tab:compilation_transformations}
\centering
\small
\setlength{\tabcolsep}{3pt}
\setlength{\fboxsep}{1pt}
\newcommand{\passcode}[1]{\colorbox{blue!3}{{\texttt{#1}}}}
\begin{tabular}{@{}p{0.28\linewidth}p{0.65\linewidth}@{}}
\hline
\textbf{Transformations} & \textbf{Purpose} \\
\hline
\passcode{--scalarize-store} & Exposes scalar accumulations from load-update-store recurrences. \\
\hline
\parbox[t]{0.3\columnwidth}{\passcode{--affine-}\\\passcode{distribute-reduction}} & Distributes polynomial reduction bodies and fissions them by $k$-varying load signatures. \\
\hline
\parbox[t]{0.3\columnwidth}{\passcode{--affine-memory-}\\\passcode{ref-canonicalization}} & Materializes dense canonical operands for transposed, packed, flattened, or sliced affine references before \mmul{} matching. \\
\hline
\multirow{3}{=}{\parbox[t]{0.3\columnwidth}{\passcode{--affine-}\\\passcode{maximize-mmul}}} & Solves dependence-preserving schedule constraints to isolate the legal \mmul{} subspace. \\
\cline{2-2}
& Checks affine accesses, dependences, and reduction semantics against the \mmul{} template. \\
\cline{2-2}
& Replaces the matched affine region with a parametric \texttt{cgra.mmul} operation for backend kernel generation. \\
\hline
\passcode{--lower-affine}$^*$ & Lowers residual affine control constructs to the SCF dialect after \mmul{} kernel extraction. \\
\hline
\passcode{--convert-scf-to-cf}$^*$ & Lowers SCF control flow to the CF dialect. \\
\hline
\multirow{2}{=}{\parbox[t]{0.3\columnwidth}{\passcode{--convert-cf-to-cgra}}} & Lowers non-\mmul{} control-flow regions to the CGRA dialect for generic CDFG mapping. \\
\cline{2-2}
& Preserves the extracted kernel operation and live-value interface in the unified CGRA program. \\
\hline
\multicolumn{2}{@{}p{0.92\linewidth}@{}}{\footnotesize $^*$Passes implemented by the MLIR infrastructure.} \\
\end{tabular}
\vspace{-1.5em}
\end{table}

\begin{figure}[t]
\centering
\small
\setlength{\fboxsep}{3pt}
\setlength{\tabcolsep}{2pt}
\newcommand{\codekw}[1]{\textcolor{blue!70!black}{\textbf{#1}}}
\newcommand{\codeop}[1]{\textcolor{violet!70!black}{#1}}
\newcommand{\codecom}[1]{\textcolor{green!45!black}{#1}}
\newcommand{\codebox}[2]{%
\fbox{\begin{minipage}[t]{0.46\textwidth}
\textbf{#1}\vspace{1pt}

{\scriptsize\ttfamily\raggedright
#2}
\end{minipage}}}
\begin{tabular}{@{}c@{\hspace{3pt}}c@{}}
\codebox{(1) Source C}{
\codekw{for} (i = 0; i < D; i++)\\
\quad \codekw{for} (j = 0; j < D; j++) \{\\
\quad\quad sum = 0;\\
\quad\quad \codekw{for} (k = 0; k < N; k++)\\
\quad\quad\quad sum += Xc[k][i] * Xc[k][j];\\
\quad\quad C[i][j] = sum;\\
\quad \}\\[2pt]
\codecom{// PCA covariance: C = Xc\^{}T * Xc}
}
&
\codebox{(2) Affine lowering}{
\codeop{affine.for} \%i = 0 \codekw{to} D \{\\
\quad \codeop{affine.for} \%j = 0 \codekw{to} D \{\\
\quad\quad \%s = 0\\
\quad\quad \codeop{affine.for} \%k = 0 \codekw{to} N \{\\
\quad\quad\quad \%a = Xc[\%k,\%i]\\
\quad\quad\quad \%b = Xc[\%k,\%j]\\
\quad\quad\quad \%s += \%a * \%b\quad\}\\

\quad\quad C[\%i,\%j] = \%s\\
\quad \}\\
\}
}
\\[-2pt]
\multicolumn{2}{c}{\scriptsize $\Downarrow$ \textit{memory-reference canonicalization}}\\[-1pt]
\codebox{(3) Canonicalized accesses}{
\codeop{affine.for} \%i = 0 \codekw{to} D \{\\
\quad \codeop{affine.for} \%j = 0 \codekw{to} D \{\\
\quad\quad \%s = 0\\
\quad\quad \codeop{affine.for} \%k = 0 \codekw{to} N \{\\
\quad\quad\quad XT[\%i,\%k] = Xc[\%k,\%i]\\
\quad\quad\quad \%s += XT[\%i,\%k] * Xc[\%k,\%j]\\
\quad\quad \}\\
\quad\quad C[\%i,\%j] = \%s\\
\quad \}\\
\}\\[2pt]
}
&
\codebox{(4) Scheduled affine form}{
\codeop{affine.for} \%i = 0 \codekw{to} D\\
\quad \codeop{affine.for} \%k = 0 \codekw{to} N\\
\quad\quad XT[\%i,\%k] = Xc[\%k,\%i]\\
\codecom{// split because $S_T(i,k) \prec S_R(i,j,k)$}\\[2pt]
\codeop{affine.for} \%i = 0 \codekw{to} D\\
\quad \codeop{affine.for} \%j = 0 \codekw{to} D\\
\quad\quad \codeop{affine.for} \%k = 0 \codekw{to} N\\
\quad\quad\quad \%s += XT[\%i,\%k] * Xc[\%k,\%j]\\
\quad\quad C[\%i,\%j] = \%s
}
\\[-2pt]
\multicolumn{2}{c}{\scriptsize $\Downarrow$ \textit{\mmul{} extraction and backend integration}}\\[-1pt]
\multicolumn{2}{c}{
\fbox{\begin{minipage}[t]{0.94\textwidth}
\textbf{(5) Integrated kernel form}\vspace{1pt}

{\scriptsize\ttfamily\raggedright
\codeop{affine.for} \%i = 0 \codekw{to} D\\
\quad \codeop{affine.for} \%k = 0 \codekw{to} N\\
\quad\quad XT[\%i,\%k] = Xc[\%k,\%i]\\[2pt]
\codeop{cgra.mmul} \{\\
\quad A = XT, B = Xc, C = C, M = D, N = D, K = N,\\
\}\\[2pt]}
\end{minipage}}}
\end{tabular}
\caption{Code-generation pipeline for partial PCA covariance code. The source uses transposed accesses, affine lowering preserves the original access functions, memory-reference canonicalization introduces a dense \texttt{XT} operand, the polyhedral scheduler splits the transpose producer from the reduction, and extraction replaces the scheduled \mmul{} subspace with a BLAS-style \texttt{cgra.mmul} .}
\Description{Five code-colored listings showing partial code for the same PCA covariance computation through source C, affine lowering, canonicalized accesses with a temporary transpose, scheduled affine form with the transpose producer split before the reduction, and integrated cgra.mmul kernel form.}
\label{fig:blas_codegen_pca}
\vspace{-1em}
\end{figure}

\vspace{-0.8em}
\subsection{Arithmetic Exposure of \mmul{} Reductions}
\label{sec: LCM}
Following the middle-end in Figure~\ref{fig: method overview}, the operation fusion scalarizes unnecessary loads and stores by storing the intermediate results in registers. It exploits the linearity of summation to perform partial computations at compile time, and fuses operations after the \mmul{} kernel.

The \texttt{--scalarize-store} pass first applies a scalar-replacement transformation targeting affine loop nests with recurrent updates to a single memory location. The pass identifies cases where an inner loop repeatedly reads a value from an invariant affine access, repeated load-store updates to the same location with a scalar recurrence, and a single final store.
The \texttt{--affine-distribute-reduction} pass then separates the canonical MAC term from surrounding arithmetic by using linearity rules. Terms with the two-load signature expected by the \mmul{} template form the candidate reduction; remaining terms stay in the residual affine code.

This transformation matches the constraint of the \mmul{} kernel: the inner loop may compute addresses, load operands, and perform MAC operations. Let $S_R(i,j,k)$ denote the reduction statement with iteration point $d=(i,j,k)^T$. Following the access-function notation in Section~\ref{sec:background}, let $M_{a_p}$ and $M_{b_q}$ denote auxiliary memory operands in the reduction body, where $M_{a_p}$ appears as a multiplicative factor of the candidate MAC term and $M_{b_q}$ appears as an additive non-MAC term. A mixed reduction can then be expressed as follows:

\begingroup
\setlength{\abovedisplayskip}{-6pt}
\setlength{\belowdisplayskip}{-2pt}
\begin{align}
    C[i,j]&=\sum_k\left(
    \prod_p M_{a_p}[F^{S_R}_{a_p}d]\cdot
    A[F^{S_R}_{A}d]\cdot B[F^{S_R}_{B}d]
    +\sum_q M_{b_q}[F^{S_R}_{b_q}d]\right) \\
          &={\alpha(i,j)}
          \sum_k A[F^{S_R}_{A}d]B[F^{S_R}_{B}d]
          {+T_k\cdot\beta(i,j)}
\end{align}
\endgroup
where $\alpha(i,j)=\prod_p M_{a_p}[F^{S_R}_{a_p}d]$, $\beta(i,j)=\sum_q M_{b_q}[F^{S_R}_{b_q}d]$, and $T_k$ is the trip count of the reduction loop. Therefore, the terms highlighted in blue can be separated from the innermost loop when their memory accesses are invariant with respect to the iterator $k$. Specifically, their affine access functions must have zero coefficients in the $k$ dimension:

\begingroup
\setlength{\abovedisplayskip}{-6pt}
\setlength{\belowdisplayskip}{-2pt}
\begin{align}
F^{S_R}_{a_p}[:,k]=0~\forall p, \quad F^{S_R}_{b_q}[:,k]=0~\forall q
\end{align}
\endgroup

Constant indices are merged and pre-computed at compile time, leaving only the necessary runtime variables in the computation chains for $\alpha$ and $\beta$. As a result, the distributed reduction form removes loop-invariant arithmetic from the kernel body and exposes a non-ambiguous accumulation variable.
Once the accumulation is rewritten, the temporary result remains in a scalar register until the reduction finishes. 
Operation fusion merges the element-wise operations performed on the accumulation result into the same computation chain, including patterns such as bias addition, scaling, and ReLU.
This removes the redundant memory round-trip, preserves the original semantics, and keeps the reduction body close to a pure MAC pattern.

\subsection{Memory-Reference Canonicalization for Hidden \mmul{}}
\label{sec:memref_canonicalization}
Memory-reference canonicalization exposes \mmul{} subspaces whose arithmetic is already a matrix product but whose memory layout is not in the form required by the kernel matcher. Let $d=(i,j,k)^T$ denote the iteration vector of a candidate reduction, the target reference pattern is

\begin{equation}
\begingroup
\setlength{\abovedisplayskip}{-4pt}
\setlength{\belowdisplayskip}{-2pt}
\label{eq:canonical_memrefs}
\begin{aligned}
F_A d &=
\begin{bmatrix}
1 & 0 & 0\\
0 & 0 & 1
\end{bmatrix}d,\quad
F_B d =
\begin{bmatrix}
0 & 0 & 1\\
0 & 1 & 0
\end{bmatrix}d,\quad
F_C d &=
\begin{bmatrix}
1 & 0 & 0\\
0 & 1 & 0
\end{bmatrix}d.
\end{aligned}
\endgroup
\end{equation}

which corresponds to the dense accesses $A[i,k]$, $B[k,j]$, and $C[i,j]$. The matcher is intentionally tied to this representation to enable extraction.

\begin{table}[t]
\caption{Memory-reference canonicalization process for hidden \mmul{} patterns.}
\label{tab:memref_canonicalization}
\centering
\small
\setlength{\tabcolsep}{4pt}
\begin{tabular}{@{}p{0.34\linewidth}p{0.54\linewidth}@{}}
\hline
\textbf{Implemented transformation} & \textbf{Example} \\
\hline
Transpose canonicalization & $B[j,k]\rightarrow B'[k,j]$ or $A[k,i]\rightarrow A'[i,k]$ \\
\hline
Affine dense packing & $M[f(i,k)]\rightarrow M'[i,k]$ or $M[g(k,j)]\rightarrow M'[k,j]$ \\
\hline
Slice-prefix preservation & $B[b,j,k]\rightarrow B'_b[k,j]$ \\
\hline
\end{tabular}
\vspace{-1em}
\end{table}

The memory-reference canonicalization pass is a pre-matching layout transformation. Its goal is to make \mmul{}-equivalent reductions visible to the fixed matcher without changing the arithmetic: when a load is equivalent to one of the target references in Equation~\eqref{eq:canonical_memrefs} but written in a different layout, the pass materializes a temporary dense view and rewrites the load to use that view.
Table~\ref{tab:memref_canonicalization} summarizes the implemented cases. Transpose canonicalization swaps the two matrix dimensions, for example rewriting $B[j,k]$ into $B'[k,j]$ so it matches the right-hand operand in Equation~\eqref{eq:canonical_memrefs}. Affine dense packing handles loads that are logically indexed by $(i,k)$ or $(k,j)$ but physically stored through a non-identity affine address. For example, if a row-major flattened operand is read as $M[W\cdot i+k]$, the pass materializes a dense view $M'[i,k]=M[W\cdot i+k]$ and rewrites the reduction to load $M'[i,k]$. Slice-prefix preservation applies the same idea when the matrix is embedded in a higher-rank tensor. The pass preserves this prefix and canonicalizes only the last two dimensions, producing a view such as $B'_b[k,j]=B[b,j,k]$. Thus each batch slice is presented to the matcher as a dense two-dimensional operand. 

All of these rewrites are conservative: if the affine access structure, static bounds, operand availability, or producer relation cannot be proven, the original program is left unchanged and the computation continues through the ordinary CDFG path.

\vspace{-0.5em}
\subsection{Dependence-Preserving \mmul{} Subspace Isolation}
The \titleacronym~implementation requires that the computation strictly conforms to the \mmul{} pattern. In this process, the \mmul{} kernel is isolated as an independent subspace while preserving the original program semantics.
This step is formulated in the polyhedral model as a dependence-preserving schedule transformation. The objective is to reshape the execution order so that the candidate \mmul{} region becomes a structurally explicit and legal kernel subspace.

For each statement within an $M$-level nested loop, its schedule matrix is defined as $\Theta \in \mathbb{R}^{(2M+1) \times (M+1)}$, where $\delta_{m,n}$ denotes the element in the $m$-th row and $n$-th column of $\Theta$. Each element satisfies $0 \leq \delta_{m,n} \leq 1$ for all $m,n \leq M$. The last column, $\delta_{m,M+1}$, encodes the relative ordering of statements within the same loop level, providing fine-grained control over intra-scope execution order. Odd-numbered rows in $\Theta$ encode loop-level ordering for loop splitting, while even-numbered rows capture the statement ordering within each loop scope. To ensure that the polyhedral transformation preserves the original program semantics and maintains a valid loop structure, the following constraints are imposed:
\begingroup
\setlength{\abovedisplayskip}{-2pt}
\setlength{\belowdisplayskip}{-2pt}
\begin{align}
    &\delta_{0,M}^i =
    \begin{cases}
        0, & i \in \mathcal{S}_M, \\
        1, & i \notin \mathcal{S}_M, \label{eq:split}
    \end{cases} \\
    \sum_i \Theta_{i,j} \leq 1, &\quad \forall j, \quad
    \sum_j \Theta_{i,j} \leq 1, \quad \forall i \bmod 2 = 1. \label{eq:row sum}
\end{align}
\endgroup
Here, $\mathcal{S}_M$ denotes the set of statements belonging to the \mmul{} kernel. Equation~\eqref{eq:split} enforces the independent extraction of \mmul{} statements for kernel isolation. Equation~\eqref{eq:row sum} ensure that each iteration variable is mapped to a unique scheduling dimension, preserving the loop hierarchy and iteration directions. Together, these constraints define a legal search space in which loop splitting can separate kernel and non-kernel computations without violating the dependence relations introduced in Section~III.
In particular, the transformation seeks a schedule in which the kernel-related statements can be grouped into a regular loop nest, while the remaining statements are kept in a separate region that can still be compiled by the general CDFG flow.

The statement ordering must ensure that the two load operations ($S_{L_0}$, $S_{L_1}$) belonging to the \mmul{} kernel are executed first, followed by the final store ($S_S$). All other statements must be scheduled afterward, regardless of the iteration indices, to avoid interference with \mmul{} execution.
\begingroup
\setlength{\abovedisplayskip}{-1pt}
\setlength{\belowdisplayskip}{-1pt}
\begin{align}
\Theta^{S_{L_0}} d_p \preceq \Theta^{S_{L_1}} d_p \prec \Theta^{S_S} d_p \prec \Theta^{S_k} d_q,  
\quad \forall k \notin \mathcal{S}_B,~ \forall d_p, d_q \in \mathcal{D}_R
\end{align}
\endgroup
where $\mathcal{S}_B$ denotes the set of statements within the \mmul{} kernel, and $\mathcal{D}_R$ is the corresponding iteration domain. 
Other statements outside the kernel can be freely reordered, as long as their memory access order remains consistent with the original program, expressed as:
\begingroup
\begin{align}
\Theta^{S_p} d_p \prec \Theta^{S_q} d_q, 
\quad \forall d_p, d_q \in \mathcal{D}_R,~ \text{if } S_p \rightsquigarrow S_q
\end{align}
\endgroup
$S_p\rightsquigarrow S_q$ denotes a dependence relation between the two statements, covering RAW, WAR, and WAW. If solutions exist for the mathematical model, \mmuls{} can be exposed and mapped to pre-compiled kernels. In the middle-end pipeline, this step follows operation fusion and prepares a legal subspace whose structure matches the target kernel template.

We use Z3~\cite{10.5555/1792734.1792766} to solve the \mmul{} that exposes schedule constraints. Since these constraints may admit multiple legal schedules, any feasible solution is sufficient for our purpose, as the goal is to expose the dominant \mmul{} region rather than optimize the residual code. When multiple \mmul{} patterns appear in the same loop nest, the formulation extracts one instance at a time; after each extracted kernel is replaced with its pre-compiled implementation, the remaining code is re-analyzed recursively until no further \mmul{} pattern can be exposed. As a result, this step produces a legal kernel-oriented loop structure for the subsequent extraction pass.

\section{Back-end Kernel Generation and Integration}
\label{sec:backend}
\subsection{Parametric Kernel Lowering and Residual Integration}
\label{sec:kernel extraction}
The back-end starts from the \texttt{cgra.mmul} operations produced by the middle-end and integrates them with the residual program. The pattern matcher requires a three-dimensional loop structure which verifies that the memory accesses can be mapped to two affine input streams and one affine output stream, corresponding to $A[i,k]$, $B[k,j]$, and $C[i,j]$. Finally, the innermost computation must be a multiply-accumulate update to the same output element, so each reduction iteration contributes one product to the accumulation for $C[i,j]$.

Candidates with non-affine accesses, ambiguous aliasing, unsupported reductions, or boundary dependences crossing the proposed kernel region are left in the ordinary CDFG path. For an accepted region, the middle-end emits a parametric \texttt{cgra.mmul} operation carrying the input and output memrefs, loop bounds, affine access descriptors, and fused-operation descriptors required by the back-end. The final \texttt{cgra.mmul} operation in Figure~\ref{fig:blas_codegen_pca} illustrates this interface for the PCA covariance kernel, while the transpose producer remains as residual code.

This strict matching policy preserves semantics. Once a region is accepted, its loop structure, affine input/output accesses, and reduction update have been checked against the supported \mmul{} form, so the accepted subspace is equivalent to the corresponding matrix product, possibly with supported fused operations. The rewrite does not relax external dependences: values live across the kernel boundary are spill/restored, and unmatched statements remain in the residual path. Thus, \texttt{cgra.mmul} changes only the implementation of the accepted subspace within the entire CDFG.

The kernel generator consumes this operation and instantiates the corresponding optimized \mmul{} assembly for the target CGRA size, as introduced in Section~\ref{sec:mmul kernel}. The remaining non-kernel code is lowered into the control-flow dialect and rewritten for CGRA execution. A CDFG is then constructed and mapped to the architecture for these residual regions. This split preserves the semantics of mixed applications: the extracted \mmul{} subspace is replaced by a pre-optimized kernel, while non-\mmul{} statements remain visible to the generic compiler.

Context generation connects the generated kernel with the surrounding basic blocks. First, kernel parameters such as base addresses and loop bounds are written into a reserved memory space, from which the kernel retrieves them during execution. Second, the dataflow across the kernel boundary is rewritten to preserve live values. The kernel occupies most of the available registers, values produced by preceding blocks cannot be assumed to remain unchanged after kernel execution. The back-end therefore performs liveness analysis, spills live-out values to memory, and restores them afterward. Finally, the kernel assembly and the residual CDFG assembly are integrated into a unified CGRA configuration; in Section~\ref{sec:exp}, the non-kernel part uses the \textit{Compigra} methodology~\cite{wang2025mlir}.

\subsection{\Mmul{} Kernel Implementation}
\label{sec:mmul kernel}
\begin{figure*}[t]
    \centering
    \begin{subfigure}[t]{0.4\textwidth}
        \centering
        \includegraphics[width=\linewidth]{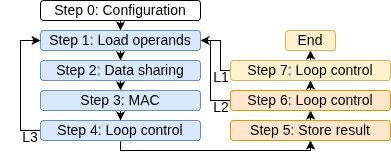}
        \caption{Kernel execution structure.}
        \label{fig:kernel_structure}
    \end{subfigure}
    \hfill
    \begin{subfigure}[t]{0.53\textwidth}
        \centering
        \includegraphics[width=0.8\linewidth]{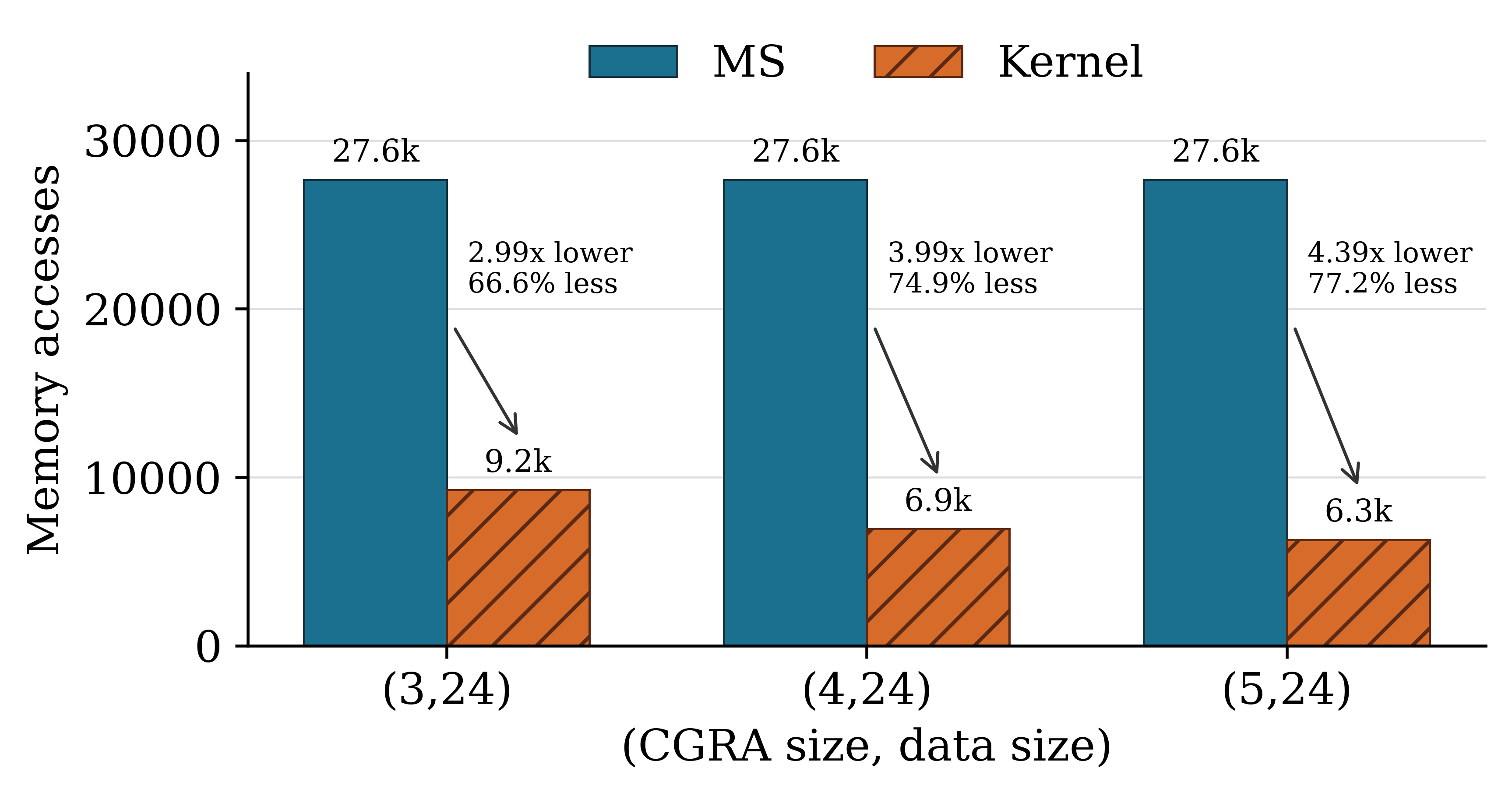}
        \caption{Memory access counts.}
        \label{fig:mmul_lwi_counts}
    \end{subfigure}
    \caption{\mmul{} kernel structure and memory-instruction reduction. The proposed kernel uses the programmability of the CGRA fabric to explicitly share loaded operands across PEs, reducing required memory accesses.}
    \vspace{-1.5em}
    \label{fig:mmul_kernel_memory}
\end{figure*}

The kernel performs matrix multiplication to compute the product $C = A \times B$, where $A \in \mathbb{R}^{N_I \times N_K}$, $B \in \mathbb{R}^{N_K \times N_J}$, and $C \in \mathbb{R}^{N_I \times N_J}$. The computation follows an output-stationary (OS) dataflow and a three-loop structure.
The optimized \mmul{} kernel is parameterizable and takes as input the base addresses and dimensions of the input matrices. The kernel operates by computing $N \times N$ tiles of the output matrix $C$, an execution strategy that naturally supports both square and rectangular matrices. The base addresses are used to determine the memory locations of the input and output data, while the matrix dimensions define the loop bounds that control the computation. These kernel parameters are stored  diagonally across the CGRA. With this layout, each row and each column contains at least one instance of the configuration information. Consequently, when a parameter is required by a given PE, it can be quickly propagated through the array using the interconnections between PEs.
The structure of the kernel implementation is illustrated in Figure~\ref{fig:kernel_structure}. This approach can be used to generate \mmul{} configurations (e.g. from the code in Figure~\ref{fig:how ms}) on time-distributed CGRAs of arbitrary size. 

\begin{figure*}[t!]
    \centering
    \includegraphics[width=\textwidth]{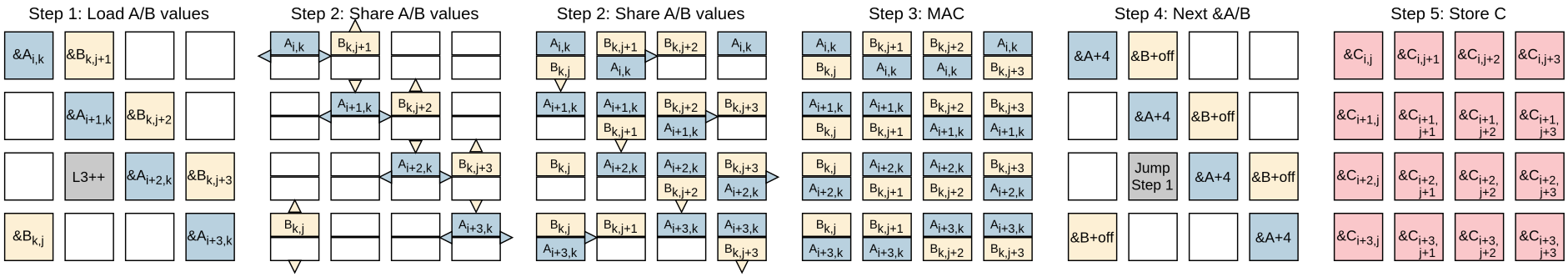}
    \caption{Mapping example of \mmul{} kernel on a $4\times4$ CGRA with torus interconnections.}
    \vspace{-1em}
    \label{fig:kernel_example_imp}
\end{figure*}

The implemented kernel exploits the programmability of CGRA's PE to enable data reuse. For a fixed \mmul{} kernel, the source-level memory references are determined by the algorithm. However, the number of external memory instructions depends on where reuse is implemented. Conventional accelerators commonly exploit reuse at the memory hierarchy level through tiling and buffering, while CDFG/MS mappings still tend to emit repeated load chains for different output elements. By exploiting CGRA programmability, our kernel explicitly shares values inside the fabric: each loaded $A[i,k]$ value is propagated across a row of PEs and each loaded $B[k,j]$ value is propagated across a column, so multiple MACs consume the same loaded operands without issuing repeated memory instructions. Figure~\ref{fig:mmul_lwi_counts} shows this reduction in memory accesses of a square 24x24 \mmul, which explains why the kernel improves performance beyond simply exposing more PE-level parallelism. An example of the mapping of the \mmul{} kernel on a $4\times4$ CGRA with torus interconnections between PEs is shown in Figure~\ref{fig:kernel_example_imp} and further discussed below.

\begin{enumerate}
\setcounter{enumi}{-1}
\item \textbf{Configuration.} All PEs load configuration values related to the affine parameters and initialize loop iterators.
\textit{Latency:} $l_{config}$.

\item \textbf{Load operands.} The PEs load the corresponding value of $A$ and $B$ from memory. The $4\times4$ example, step 1, shows that it loads 4 values from $A$ and $B$ simultaneously. \textit{Latency:} $l_{ld}$.

\item \textbf{Data sharing.} Values of $A$ are shared across rows and values of $B$ across columns. The loaded values are propagated to compute 16 elements for $4\times4$ CGRAs, as shown in step 2 in Figure~\ref{fig:kernel_example_imp}.
\textit{Latency:} $l_{sh}$ depends on the CGRA size and the interconnections between PEs.

\item \textbf{MAC.} Multiply-accumulate operation.
\textit{Latency:} $l_{mac}$.

\item \textbf{Loop control (L3).} Addresses of $A$ and $B$ are updated and execution returns to Step~1. The address computation is reduced to offset-only calculations due to the decomposition being handled at compile time, thereby reducing runtime overhead, as shown in Step 4 of Figure~\ref{fig:kernel_example_imp}.
For \(N<4\), higher register pressure requires incrementing in a single PE and sharing the result across pointer PEs, resulting in the additional 1 CC overhead.
\textit{Latency:} $l^{\text{ctrl}}_{L3}=\left\{1(N\geq4),2(N<4)\right\}$.

\item \textbf{Store result.} The computed value is written in memory. This implies sharing the store address across the required cells while being updated for the next iteration. 
\textit{Latency:} $l_{sh} + l_{st}$.

\item \textbf{Loop control (L2).} Addresses of $A$ and $B$ are updated for the next iteration of loop $L2$ (column tiling). The counter of loop $L3$ is reset and execution jumps to Step~1. 
\textit{Latency:} $l^{\text{ctrl}}_{L2}$.

\item \textbf{Loop control (L1).} Addresses of $A$ and $B$ are updated for the next iteration of loop $L1$ (row tiling). The counter of loop $L2$ is reset and execution jumps to Step~1. 
\textit{Latency:} $l^{\text{ctrl}}_{L1}$.
\end{enumerate}

The  number of execution cycles of the kernel, considering the parallelism achievable in an $N \times N$ CGRA, can be expressed in closed form as: 

\begingroup
\setlength{\abovedisplayskip}{-4pt}
\setlength{\belowdisplayskip}{-2pt}
\[
\begin{aligned}
\Bigg[ \Big( 
    &(l_{ld} + l_{sh} + l_{mac} + l_{L3}^{\text{ctrl}}) N_K + l_{sh} + l_{st} + l^{\text{ctrl}}_{L2}
\Big) \frac{N_J}{N}
+ l^{\text{ctrl}}_{L1} \Bigg]
\frac{N_I}{N}
\end{aligned}
\]
\endgroup

Step 0 is excluded from this formulation because it is executed only once and its cost is negligible compared to the total number of execution cycles. 

Compared to the MS approach, this formulation achieves significant speedups through a combination of tiling and data sharing. On the one hand, tiling reduces the number of iterations of the outer loops from $N_I$ and $N_J$ to $N_I/N$ and $N_J/N$, respectively. On the other hand, the innermost loop leverages data sharing to reduce the number of memory accesses, at the cost of additional instructions. Comprehensive results evaluating runtime efficiency are presented in Section~\ref{sec:exp}.

When the output matrix dimensions ($N_I, N_J$) are not exact multiples of the CGRA spatial dimension $N$, the matrix product results in irregular, partial boundary tiles. Unrolling or parallelizing these partial tiles within the optimized kernel would reintroduce dimension mismatch issues. To handle non-divisible dimensions, our framework partitions the problem: the maximum possible number of full $N \times N$ tiles is processed by the optimized \mmul{} kernel, while any remaining partial tile is compiled separately using the standard compiler backend via its corresponding CDFG. Accordingly, the analytical execution-cycle model in Section 6.2 applies to the complete tiles, with the tile-count factors $N_I/N$ and $N_J/N$ replaced by $\lfloor N_I/N \rfloor$ and $\lfloor N_J/N \rfloor$, respectively.

Finally, assembly generation materializes the steps above as a pre-compiled instruction region.
Each instruction specifies the operation and register usage of one PE, with multiple instructions executing simultaneously in the same clock cycle. The generated region therefore preserves the structure of the kernel implementation: configuration, operand loading, data sharing, accumulation, result storage, and loop control are emitted in the same order as described above. For each target CGRA size, the backend emits the corresponding assembly, fills in the program-specific parameters, and inserts the resulting region into the final CGRA configuration together with the residual code generated by \textit{Compigra}.
\section{Experiments}
\label{sec:exp}
\subsection{Setup}
\label{subsec:setup}

\subsubsection{Target CGRA Platform}

\begin{figure*}[t]
    \centering
    \begin{subfigure}[b]{0.34\textwidth}
        \centering
        \includegraphics[width=\linewidth]{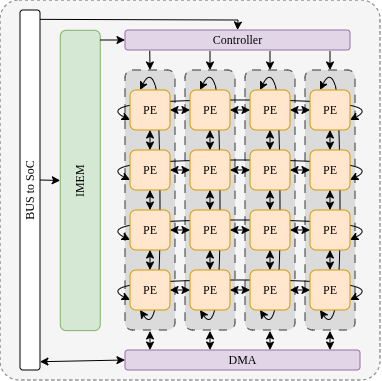}
        \caption{Target CGRA architecture.}
        \label{fig:cgra_abstract}
    \end{subfigure}
    \hfill
    \begin{subfigure}[b]{0.6\textwidth}
        \centering
        \includegraphics[width=0.9\linewidth]{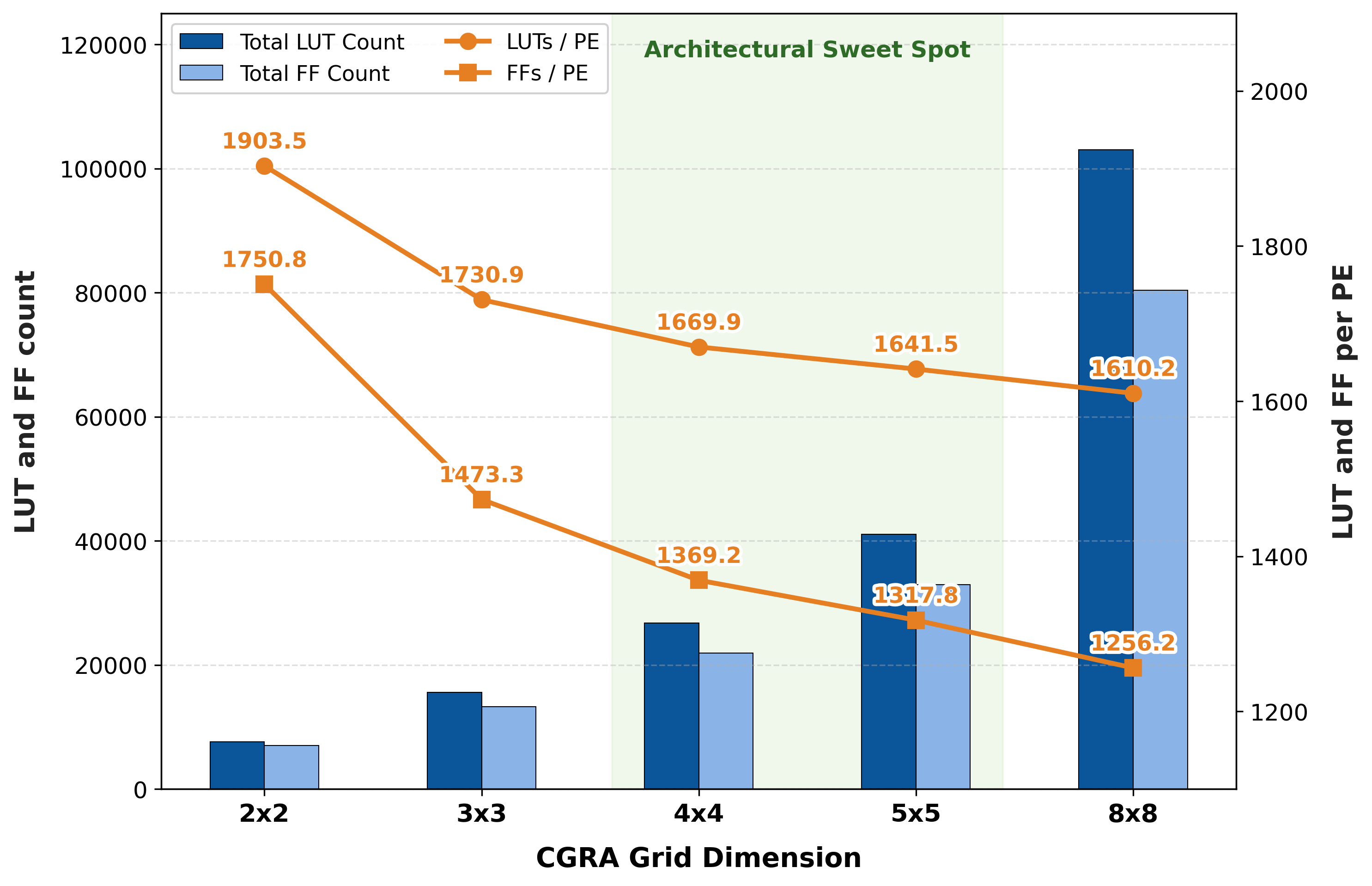}
        \caption{{Area cost scaling and normalized cost per PE.}}
        \label{fig:cgra_scaling}
    \end{subfigure}
    \caption{Target CGRA platform overview: (a) CGRA architecture scheme, and (b) area scaling analysis for different OpenEdgeCGRA grid dimensions.}
    \vspace{-1.5em}
    \label{fig:cgra_platform}
\end{figure*}

We evaluate the proposed framework on a two-dimensional CGRA abstraction with a torus network (Figure~\ref{fig:cgra_abstract}), though our flow seamlessly targets arbitrary interconnect topologies, including diagonal links or shared memory. 

{For runtime execution, the generic schedule generated by our framework is tailored to the specific Instruction Set Architecture (ISA) of the OpenEdgeCGRA platform~\cite{openEdgeCGRA}, which implements the topology in Figure~\ref{fig:cgra_abstract}. OpenEdgeCGRA is a time-distributed CGRA integrated within the X-HEEP SoC~\cite{machetti2025xheep}, where a CV32E40P RISC-V host core orchestrates execution. Table~\ref{tab:cgra_architecture} summarizes the full architectural specifications and execution parameters.}

\begin{table}[h!]
\centering
\caption{Detailed architectural and execution parameters of the target CGRA platform.}
\label{tab:cgra_architecture}
\small
\begin{tabular}{ll}
\toprule
\textbf{Architectural Parameter} & \textbf{Specification / Value} \\
\midrule
Grid Size                    & ($3\times3$), ($4\times4$), ($5\times5$) \\
PE Architecture              & 32-bit ALU with local 4-entry Register File \\
Instruction Memory (IMEM)    & 128 entries (32-bit width) per PE \\
Operand Widths               & 32-bit (data and instructions) \\
Interconnect                 & 2D Torus (nearest neighbour connections with wrap-arounds) \\
Memory Access \& Interconnect & 1 DMA per CGRA column accessing system memory via bus \\
Execution Model              & Synchronized PC execution across all PEs \\
Operations Latency  & ALU/Logic: 1 cycle, Mult: 3 cycles, Load: 2 cycles, Store:  1 cycle \\ 
\bottomrule
\end{tabular}
\end{table}

To select representative grid dimensions for our evaluation, we conducted an area scaling analysis across various CGRA dimensions using AMD/Xilinx Vivado~\cite{xilinx_vivado} for FPGA synthesis, as shown in Figure~\ref{fig:cgra_scaling}. While total logic resource usage grows rapidly with grid size, configurable logic absorbs the non-linear global interconnect and control overhead. Normalizing resource utilization per PE reveals an architectural sweet spot around $4\times4$ and $5\times5$ configurations. In smaller grids (e.g., $2\times2$), top-level control and memory interfaces account for a high fraction of the total logic, resulting in $1,903.5\text{ Look-Up Tables (LUTs)/PE}$ and $1,750.8\text{ Flip-Flops (FFs)/PE}$. As grid dimensions scale to $3\times3$, $4\times4$, and $5\times5$, these global overheads are effectively amortized, causing logic density to drop sharply and stabilize around $1,641.5\text{--}1,669.9\text{ LUTs/PE}$ and $1,317.8\text{--}1,369.2\text{ FFs/PE}$. Beyond $5\times5$ (e.g., $8\times8$), marginal logic efficiency gains per PE diminish significantly ($1,610.2\text{ LUTs/PE}$ and $1,256.2\text{ FFs/PE}$) while resource requirements increases notably. Therefore, we select $3\times3$, $4\times4$, and $5\times5$ configurations for our experimental evaluation, as they capture both the scaling transition and the optimal region of area-efficient CGRA design.

\subsubsection{Benchmarks}

The evaluation employs a representative set of benchmarks that exercise both canonical dense linear-algebra kernels and application-level computations with integrated or hidden \mmul{} structure. The suite includes standard PolyBench kernel, \textit{gemm}, \textit{2mm}, and \textit{3mm}~\cite{pouchet2012polybench}; matrix-multiplication variants, \textit{DenseLinear}, \textit{BatchedLinear}, \textit{MatMul-ReLU}, and \textit{QuantizedMatMul}; and application-level kernels, \textit{PCA}~\cite{mackiewicz1993pca}, \textit{KalmanPredict}, and \textit{Kalman-Cov}~\cite{welch1995kalman}. 

Table~\ref{tab:blas-benchmark-transform} summarizes the benchmark characteristics and the compiler transformations required to expose and integrate reusable \mmul{} kernels. A checkmark indicates that the corresponding pass is necessary for successful extraction in that benchmark: disabling the pass prevents at least one \mmul{} region from being emitted as a \texttt{cgra.mmul} operation. The selected benchmarks cover dense matrix multiplication, chained linear algebra operations, fused element-wise operations, quantization-related post-processing, and data pre-processing steps involving transposition, accumulation, and covariance computations.
We consider matrix dimensions of $24\times24$ and $60\times60$ for the full benchmark evaluation, with the latter resulting in memory footprints exceeding 100~KB for selected benchmarks. We further include an isolated \mmul{} kernel case study with irregular matrix sizes to evaluate kernel usage and residual-region overhead under non-divisible dimensions.

For all evaluated benchmark configurations, functional correctness was validated by verifying that the execution outputs produced by the CGRA match the golden static reference outputs identically in every test case.

\begin{table*}[t]
\centering
\caption{{Benchmark transformation coverage across evaluated workloads.}}
\label{tab:blas-benchmark-transform}
\small
\begin{tabularx}{\textwidth}{l X cccccc}
\toprule
\textbf{Benchmark} & \textbf{Attributes} & \textbf{Sc} & \textbf{Dr} & \textbf{Mr} & \textbf{Mm} & \textbf{Ctx} & \textbf{Fo} \\
\midrule
\textit{gemm} & Adds output scaling and accumulation & \checkmark & \checkmark & -- & \checkmark & \checkmark & -- \\
\textit{2mm} & Chains MMULs and scalar coefficients across dependent products & \checkmark & \checkmark & -- & \checkmark & \checkmark & \checkmark \\
\textit{3mm} & Three chained \mmul{} regions with intermediate matrices & \checkmark & -- & -- & \checkmark & \checkmark & -- \\
\textit{DenseLinear} & -- & \checkmark & -- & -- & \checkmark & -- & -- \\
\textit{BatchedLinear} & Batched dimension with rank-3 tensors & \checkmark & -- & -- & \checkmark & \checkmark & -- \\
\textit{MatMul-ReLU} & Post-\mmul{} ReLU/clamp & \checkmark & -- & -- & \checkmark & \checkmark & \checkmark \\
\textit{QuantizedMatMul} & Intertwines the \mmul{} with per-row bias and per-column scale terms & \checkmark & \checkmark & -- & \checkmark & \checkmark & -- \\
\textit{PCA} & Uses centered data and computes covariance & \checkmark & -- & \checkmark & \checkmark & \checkmark & -- \\
\textit{Kalman-Predict} & Combines matrix-vector prediction with a following matrix-matrix update & \checkmark & -- & -- & \checkmark & \checkmark & -- \\
\textit{Kalman-Cov} & Transpose-style \mmul{} access and additive noise & \checkmark & \checkmark & \checkmark & \checkmark & \checkmark & -- \\
\bottomrule
\end{tabularx}

\vspace{0.4em}
\begin{minipage}{\textwidth}
\centering
\scriptsize
\textbf{Sc}: scalarize-store; \quad
\textbf{Dr}: reduction restructuring; \quad
\textbf{Mr}: memory/layout rewrite; \quad
\textbf{Mm}: \mmul{} exploitation; \quad
\textbf{Ctx}: non-MMUL context; \quad
\textbf{Fo}: fused output op.
\end{minipage}
\vspace{-1.5em}
\end{table*}

\vspace{-0.5em}
\subsubsection{Baselines}
For CGRA-based comparison, we evaluate our approach against an alternative strategy employing \textit{Compigra}~\cite{wang2025mlir} for mapping the entirety of applications  which, to the best of our knowledge, represents the state of the art of compilers performing CDFG mapping for complex applications on CGRAs. 
The baseline applies two optimization techniques: (1) MS for innermost loops, and (2) loop unrolling along the column dimension of the output array.
The unrolling factor ($U$) is set to $\lfloor N^{2} / 2 \rfloor$, where $N$ is the CGRA size.  The unrolling factor is selected to achieve the maximum compute ability supported by the available hardware resources. Under this configuration, two PEs form a group that enables simultaneous loads from both input matrices. Within each group, the computations of $C[i][j+u]$ for $u = 1,\ldots,U$ are executed in parallel.

For a broader runtime evaluation, we also compare against two other edge accelerators. These baselines are selected to cover common edge platforms and specialized \mmul{} accelerators: e-GPU~\cite{machetti2025egpu} is a lightweight multi-threaded GPU designed for edge systems, while systolic-array-like and mixed architectures are widely used in CGRA-related accelerator designs, as discussed in Section~\ref{subsec:cgra-compilation}. We also compare with a 12$\times$12 systolic array (SA)~\cite{systolic_array} coupled with the X-HEEP SoC~\cite{machetti2025xheep}.\footnote{The SA size is selected to achieve a synthesized area comparable to the considered OpenEdgeCGRA instance (0.4 mm$^2$ in TSMC 65nm) for fair comparison.} 
The SA+CPU baseline uses the same front-end and middle-end transformation pipeline as Kernel-CGRA. Therefore, both approaches expose the same affine \mmul{} regions and replace them with explicit kernel invocations. The difference is in the back-end lowering and residual-code execution: Kernel-CGRA lowers each extracted \mmul{} operation to the proposed CGRA kernel and maps residual non-\mmul{} regions through the CDFG CGRA path, whereas SA+CPU lowers each extracted \mmul{} operation to the systolic-array accelerator and executes residual non-\mmul{} regions on the host CPU. All accelerators are integrated with the same X-HEEP SoC, described in Section~\ref{subsec:setup}.

\subsection{Code generation}

\begin{figure*}[t]
    \centering
    \includegraphics[width=0.95\textwidth]{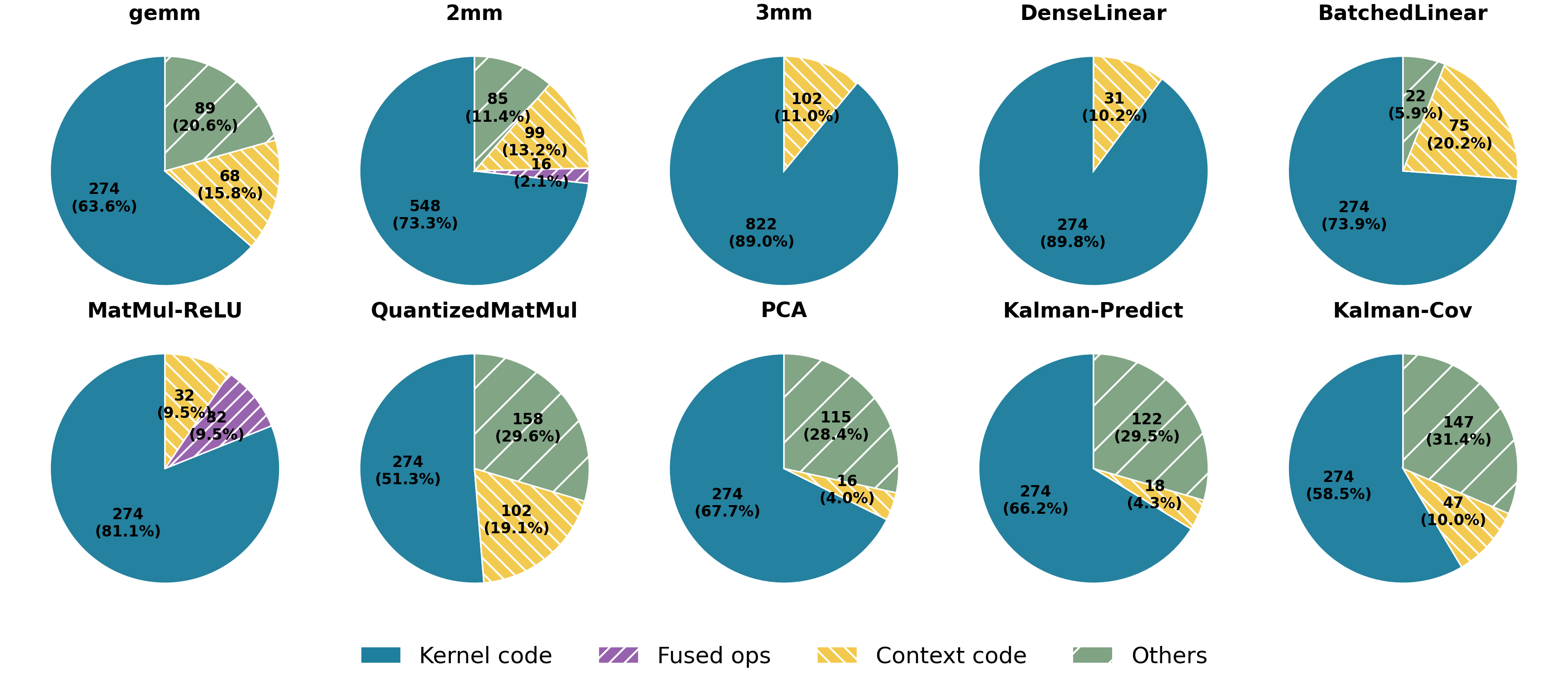}
    \caption{Code composition after decomposition and kernel extraction. Each pie reports the number and percentage of operations emitted as pre-compiled kernel code, fused operations, context code for preserving program semantics across kernel boundaries, and residual non-\mmul{} code.}
    \label{fig:decompose_pies}
    \vspace{-1em}
\end{figure*}

Figure~\ref{fig:decompose_pies} reports instruction counts for the generated-code composition for the $4\times4$ CGRA configuration, where a standard \mmul{} kernel instance is implemented with 273 instructions. The kernel-code slice corresponds to instructions replaced by pre-compiled \mmul{} invocations. Context-generation code includes the instructions needed to prepare kernel operands, preserve live values, and reconnect the extracted kernel with the surrounding CDFG. Fused operations are element-wise operations that can be integrated into the \mmul{} scope to reduce memory accesses. The remaining \textit{Others} slice contains residual non-\mmul{} operations for auxiliary functionality, which are still mapped through the regular CDFG flow.

The decomposition behavior follows the benchmark structure. Pure or repeated matrix multiplication workloads, such as \textit{DenseLinear} and \textit{3mm}, have large kernel-code shares because most operations belong to reusable \mmul{} regions. \textit{BatchedLinear} includes more context-generation code, since each \mmul{} invocation inside the batch loop requires live data to be restored, retrieved, and passed between iterations. \textit{MatMul-ReLU} and \textit{2mm} contain fused operations that are integrated into the \mmul{} code region. \textit{gemm} and \textit{2mm} involve additional arithmetic operations beyond the pure \mmul{} scope, and \textit{QuantizedMatMul} requires additional scaling and bias.

\begin{figure*}[t]
    \centering
    \includegraphics[width=0.95\textwidth]{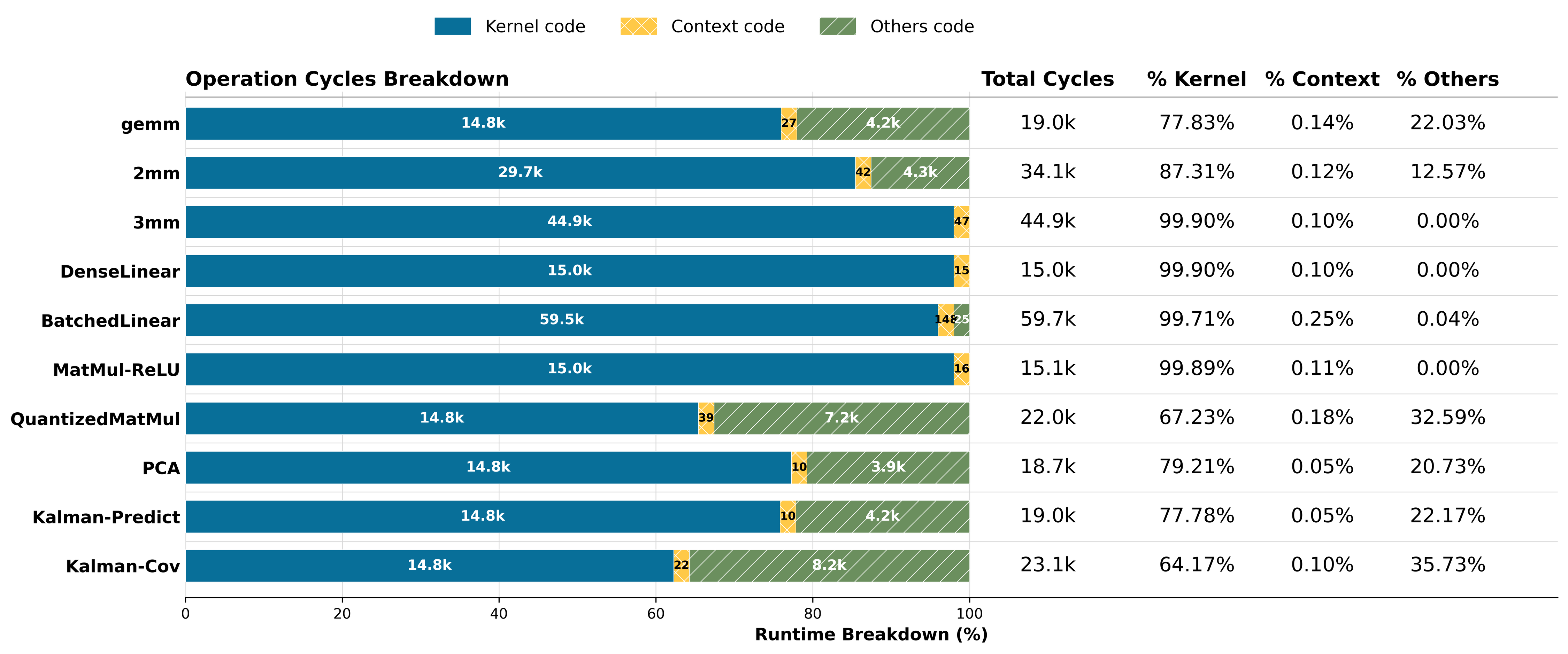}
    \caption{Runtime cycle decomposition of the proposed flow on a $4\times4$ CGRA with $N_{IJK}=24$. Cycles are grouped into pre-compiled kernel execution, context code, and residual non-\mmul{} code.}
    \label{fig:cycle_decomposition}
\end{figure*}

Application-level kernels exhibit a larger residual component, as their computation is beyond a single dense matrix product pattern. In \textit{PCA}, \textit{KalmanPredict}, and \textit{KalmanCovUpdate}, the \textit{Others} portion accounts for 28.4\%, 29.5\%, and 31.4\% of the generated code, respectively, due to transposition, covariance accumulation, and update logic around the extracted \mmul{} regions. It indicates that the proposed decomposition is effective beyond standalone \mmul{} kernels while still leaving the non-\mmul{} parts compiled for CGRA execution.

\subsection{Runtime cycle decomposition}

Figure~\ref{fig:cycle_decomposition} reports the runtime contribution of each generated-code category for a $4\times4$ CGRA and $N_{IJK}=24$. Fused operations are integrated into the kernel and are therefore not visible as a separate category in this cycle breakdown. In addition, the context-code and \textit{Others} segments are visually emphasized when their actual proportional width would be difficult to discern; their labels report the corresponding cycle counts. Although context-generation code can occupy a non-negligible fraction of the static code space in Figure~\ref{fig:decompose_pies}, its runtime impact is limited. Across all benchmarks, context code accounts for at most 0.25\% of total cycles, with absolute costs ranging from 10 to 148 cycles. The instructions used to prepare operands and reconnect extracted kernels are mostly fixed overheads that are effectively amortized by the execution of the pre-compiled \mmul{} kernel.

The remaining non-\mmul{} code has a stronger dependence on benchmark structure. For pure or near-pure matrix multiplication workloads, such as \textit{DenseLinear}, \textit{MatMul-ReLU}, and \textit{3mm}, the kernel accounts for at least 99.89\% of runtime, and residual code is negligible. In contrast, \textit{gemm}, \textit{PCA}, \textit{KalmanPredict}, \textit{QuantizedMatMul}, and \textit{KalmanCovUpdate} spend 20.73\%--35.73\% of their cycles in \textit{Others}. These cycles mainly correspond to computations detached from the intertwined \mmul{} region, such as scaling, quantization, transposition, covariance updates, or other pre- and post-processing steps. In these mixed workloads, the kernel remains the dominant component, contributing 64.17\%--79.21\% of total runtime.

\subsection{Compilation time}

\begin{figure*}[t]
    \centering
    \includegraphics[width=0.95\textwidth]{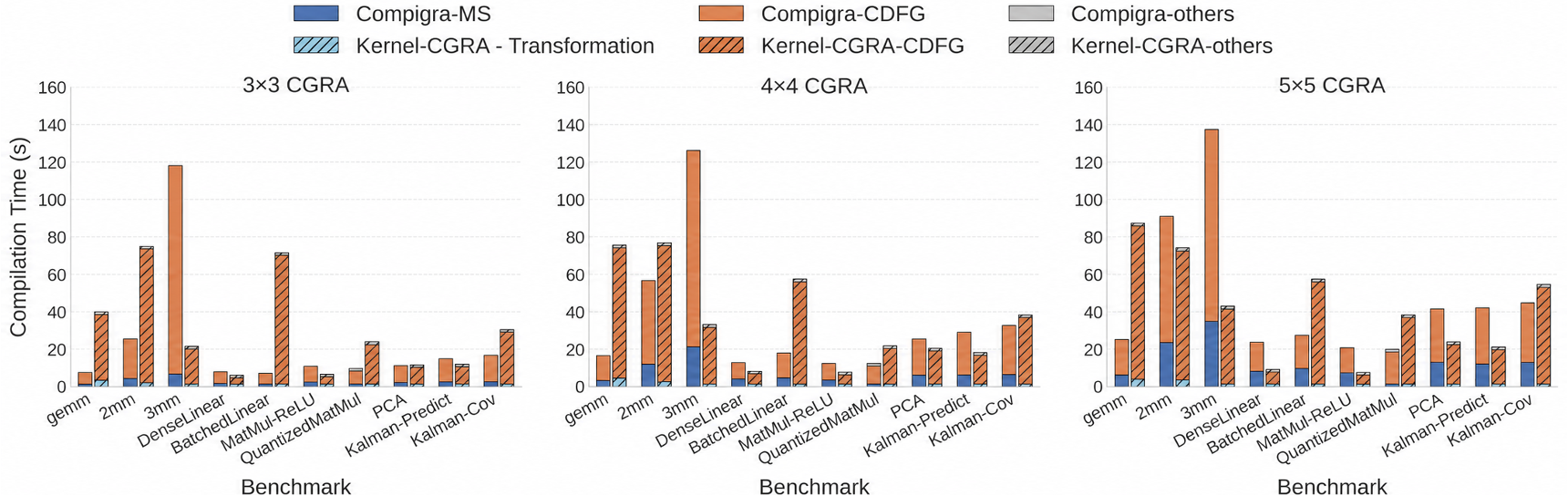}
    \caption{Compilation time of benchmark applications for CGRAs with various size using the proposed kernel pre-compilation approach, compared with \textit{Compigra} with MS optimization \cite{wang2025mlir}.}
    \label{fig:compilation time}
    \vspace{-1.5em}
\end{figure*}

Figure \ref{fig:compilation time} compares the compilation time of the proposed kernel pre-compilation flow with \textit{Compigra} employing MS across three CGRA configurations: 3$\times$3, 4$\times$4, and 5$\times$5. 
Overall, our approach does not increase the front-end compilation cost for benchmarks with strong \mmul{} structures. In several cases, compilation time is reduced by reusing precompiled kernels, thereby significantly reducing the search space in the transformation and mapping stages. This effect is most pronounced in \textit{DenseLinear}, \textit{MatMul-ReLU}, and \textit{3mm}, where matrix multiplication dominates the workload.

The stage-wise breakdown in Figure \ref{fig:compilation time} shows that the impact is not uniform across compilation stages. The specific \mmul{} kernel is generated offline and cached, while application-dependent parameters are specialized during compilation with only millisecond-level overhead, negligible compared with CDFG scheduling and mapping.
In the proposed flow, the transformation stage is typically shorter and sometimes negligible, while CDFG generation and mapping still incur overhead to handle residual non-\mmul{} regions and connect them with pre-compiled kernels. This effect is visible in \textit{gemm}, \textit{2mm}, and \textit{Kalman-Cov}, where surrounding computations limit the compilation-time reduction. \textit{BatchedLinear} is a notable case: \textit{Compigra} compiles it relatively quickly because the outer loop mainly adds control overhead, whereas our flow spends additional effort generating the context needed to reuse the inner \mmul{} kernel.

Across all CGRA configurations, the same trend is observed: increasing the CGRA size raises the absolute compilation time for both approaches. Despite this increase, the proposed method consistently provides the largest benefit for benchmarks dominated by reusable \mmul{} kernels. These results indicate that kernel pre-compilation is particularly effective for regular linear-algebra workloads, while maintaining reasonable compilation time for applications with intertwined \mmul{}s.

\subsection{Runtime efficiency}
\label{sec: runtime}

\begin{figure*}[ht]
    \centering
    \includegraphics[width=\textwidth]{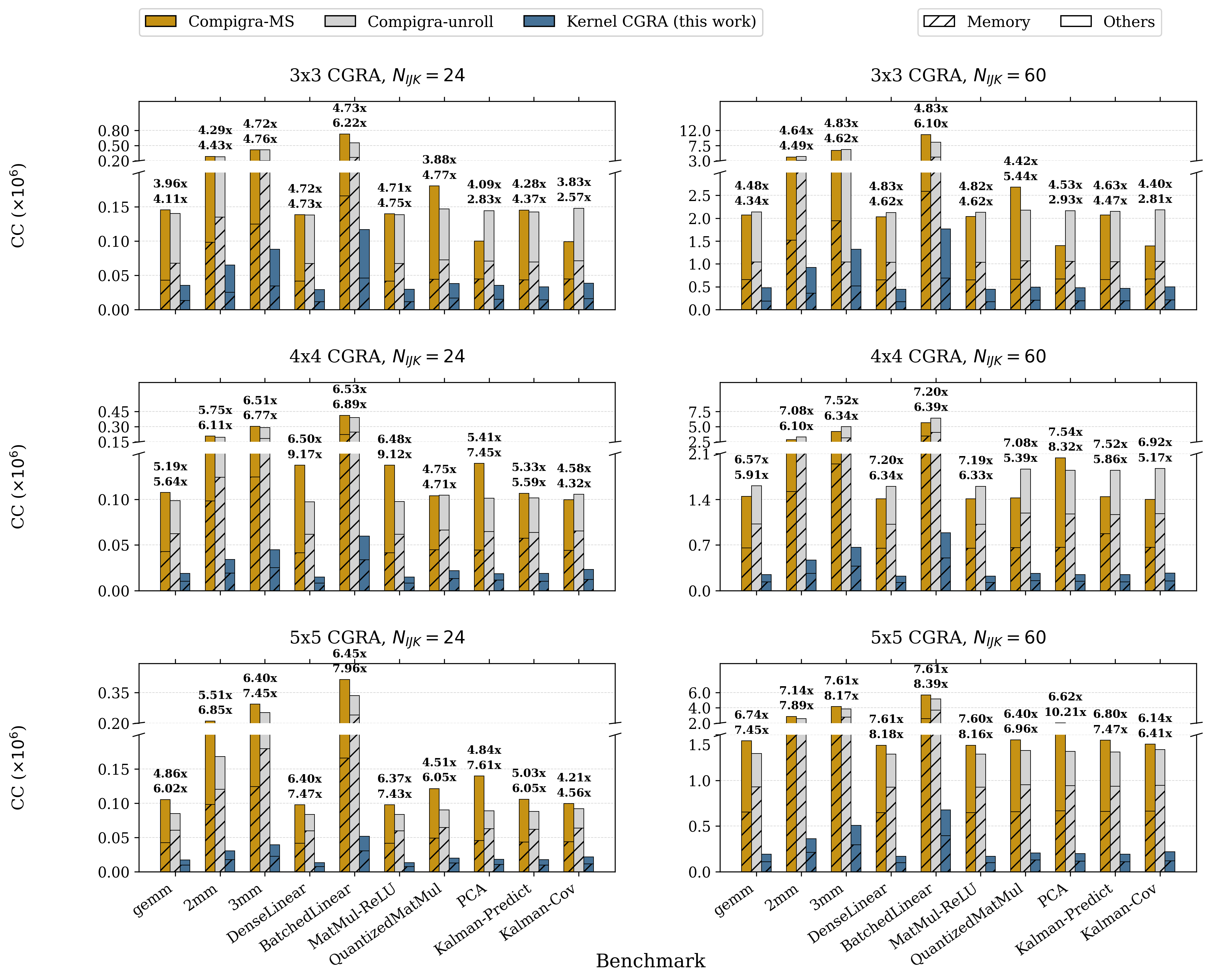}
    \caption{{Execution cycle count (CC) comparison between \textit{Compigra} and the proposed strategy with pre-compiled \mmul kernels, across CGRA configurations and matrix sizes. The numbers annotated above the clipped bars indicate the speedup of the proposed approach w.r.t. MS and unroll compilation (top and bottom values, respectively), with a broken y-axis used to improve readability, and hatched regions represent cycles spent on memory transactions.}}
    \vspace{-1em}
    \label{fig:hw_results_plots}
\end{figure*}

Figure~\ref{fig:hw_results_plots} presents the execution-cycle counts compiled through our approach compared to the baselines. The figure shows that the proposed kernel-based flow consistently outperforms both \textit{Compigra}-MS and \textit{Compigra}-unroll for every benchmark, matrix size, and CGRA configuration. Overall, the runtime speedup over fully compiler-generated code ranges from $2.5\times$ to $10.2\times$. 

As expected, the largest gains are observed for benchmarks that most closely match the \mmul{} kernel structure, such as \textit{DenseLinear} and \textit{MatMul-ReLU}. Benchmarks with repeated \mmul{} patterns, such as \textit{BatchedLinear} and \textit{3mm}, also achieve significant improvements due to the reuse of the optimized kernel across multiple invocations. Moreover, even 
benchmarks such as \textit{gemm}, \textit{PCA}, and the two \textit{Kalman} stages, for which a larger fraction of their execution is non-\mmul{}, still exhibit tangible speedups. The broken y-axis and the annotated values further emphasize the large absolute runtime gap for the heaviest workloads.

{To validate the mechanism driving these gains, Figure~\ref{fig:hw_results_plots} isolates the memory cycle breakdown (hatched regions). While baseline approaches spend a dominant fraction of their execution cycles stalled on memory transactions due to repeated load chains, our kernel reduces total memory cycles. Specifically, memory-bound cycles are reduced by $2.6\times$ to $7.1\times$ compared to \textit{Compigra} using MS, and by $2.0\times$ to $9.3\times$ compared to compiler compilation with unrolling. By exploiting the CGRA fabric to share loaded $A$ and $B$ operands across rows and columns of PEs, the kernel eliminates redundant load instructions, ensuring memory overhead does not saturate available bandwidth.}

The impact of matrix size on performance differs across methods. When the matrix size increases from $N_{IJK}=24$ to $N_{IJK}=60$, execution cycles increase across all methods, but the gap between the proposed approach and the baselines widens. Larger matrices expose more parallelism to be exploited, and the fixed control and kernel-integration overheads are further amortized over a much larger amount of computation. {Furthermore, the reduction in memory transactions scales favorably with matrix size, as the relative overhead of initial operand loads diminishes compared to the volume of compute operations.} This improves the effectiveness of the pre-compiled kernel and increases data reuse across repeated MAC operations. As a result, the speedup tends to increase slightly with matrix size, particularly in benchmarks that deviate from the ideal \mmul{} pattern.

The larger CGRA size enables larger opportunities for data sharing and parallelism exploitation.
Moving from a $3\times3$ to a $5\times5$ array reduces execution cycles for all approaches, but the reduction is substantially larger for the kernel CGRA implementation, since the pre-compiled kernel can directly exploit the additional PEs by distributing independent output-element computations across a larger array. Crucially, as the array grows, our spatial data-sharing strategy allows memory cycles to remain low despite the increase in parallel compute operations. In contrast, \textit{Compigra} remains limited by dependence-constrained MS and by the achieved Initiation Intervals (IIs). IIs equal to 3, 2 and 2 are observed for CGRA sizes $3\times3$, $4\times4$ and $5\times5$, respectively.  II values would not decrease further with larger CGRA sizes, since the available data parallelism in the inner loops is already saturated. As a result, the benefit from additional hardware diminishes. However, the kernel-based flow can continue to leverage the additional resources to further reduce execution time.

Overall, Figure~\ref{fig:hw_results_plots} indicates that the proposed strategy provides both high speedup and better scalability. Larger data sizes increase the amount of reusable regular computation, and larger CGRAs provide more parallel resources to exploit that structure.

\subsection{Kernel Usage Across Matrix Sizes}
\label{subsec:kernel-usage-matrix-sizes}

\begin{figure*}[t]
    \centering
    \includegraphics[width=\textwidth]{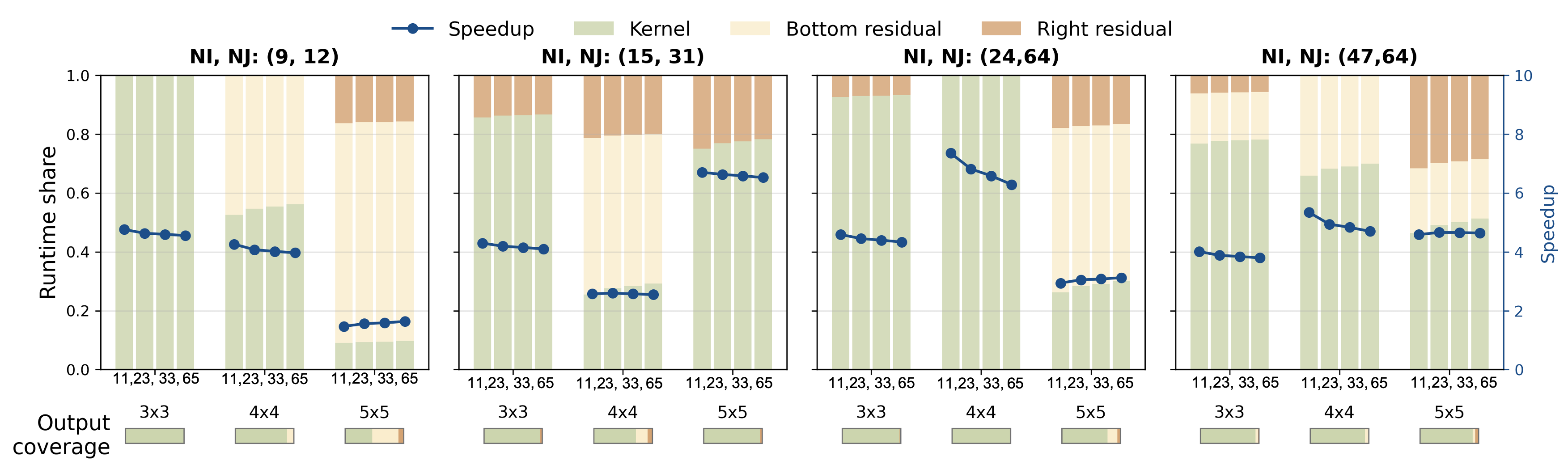}
    \caption{Runtime decomposition and speedup of the proposed \mmul{} kernel across irregular matrix sizes and CGRA configurations. Bars represent the runtime contribution of the kernel and the residual bottom and right regions. Lines indicate speedups relative to the CDFG baseline, while the output coverage bar displays the distribution of the output elements across these regions.}\
    \vspace{-2em}
    \label{fig:mmul_runtime_breakdown}
\end{figure*}

\begin{figure*}[t]
    \centering
    \includegraphics[width=\textwidth]{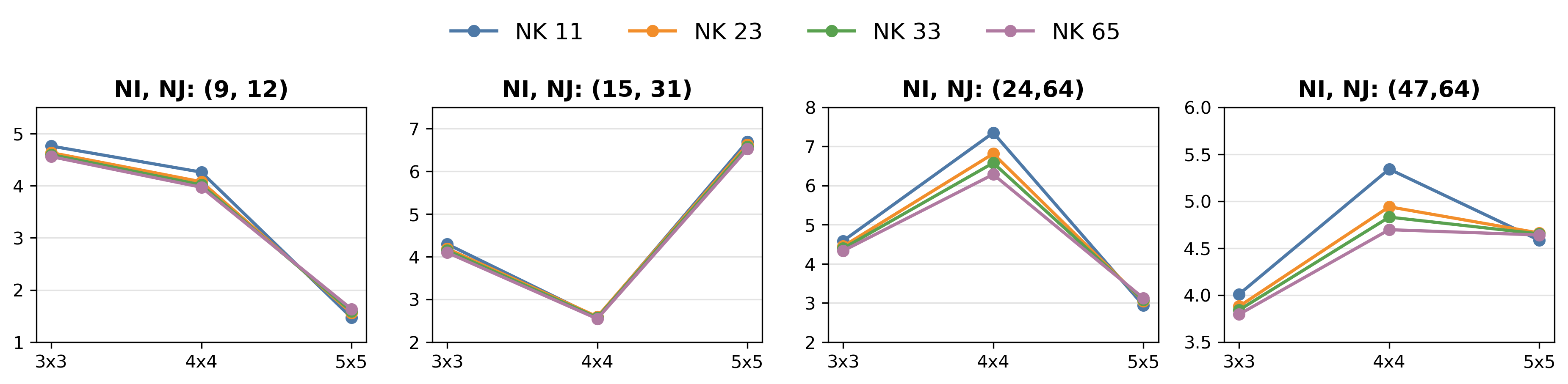}
    \caption{Speedup over CDFG baseline of the proposed \mmul{} kernel across irregular output dimensions and reduction sizes.}
    \vspace{-1em}
    \label{fig:mmul_speedup_trend}
\end{figure*}

We study the effect of irregular matrix dimensions on the kernel performance, since the residual non-kernel regions compiled via the CDFG path are not directly determined by the selected \mmul{} kernel tile size. This case study evaluates how residual regions not covered by the kernel impact the overall execution when the output dimensions are not divisible by the CGRA size.

Figure~\ref{fig:mmul_runtime_breakdown} reports the runtime decomposition for irregular output sizes on $3\times3$, $4\times4$, and $5\times5$ CGRAs, with $N_K \in \{11,23,33,65\}$. Although the \mmul{} kernel accounts for the majority of output elements, small residual regions can still introduce notable runtime overhead--particularly for small $N_I$ and $N_J$ values-because CDFG--mapped residuals fail to fully exploit array density. As $N_K$ increases, the modulo scheduler exposes greater loop-level parallelism, improving speedup until it plateaus. Larger CGRAs can amplify these residual overheads because an $N\times N$ array can yield larger remainders (up to $N-1$). Nevertheless, when residual coverage remains small, the high compute density of the main kernel dominates overall performance.

Figure~\ref{fig:mmul_speedup_trend} depicts the speedup trends across various CGRA sizes for each irregular output shape. Residual regions disproportionately impair speedup for small output dimensions; for example, at $(N_I,N_J)=(9,12)$, a $5\times5$ array loses execution efficiency due to the relatively large residual workload. As the output dimensions scale to $(47,64)$, this residual overhead is amortized, mitigating its performance penalty. Notably, the $4\times4$ array achieves optimal scaling for $(24,64)$ because both dimensions align perfectly with the array size, simultaneously eliminating residual overhead and maximizing parallelism.

\subsection{Comparison with other accelerators}

\begin{figure}
    \centering
    \includegraphics[width=0.98\textwidth]{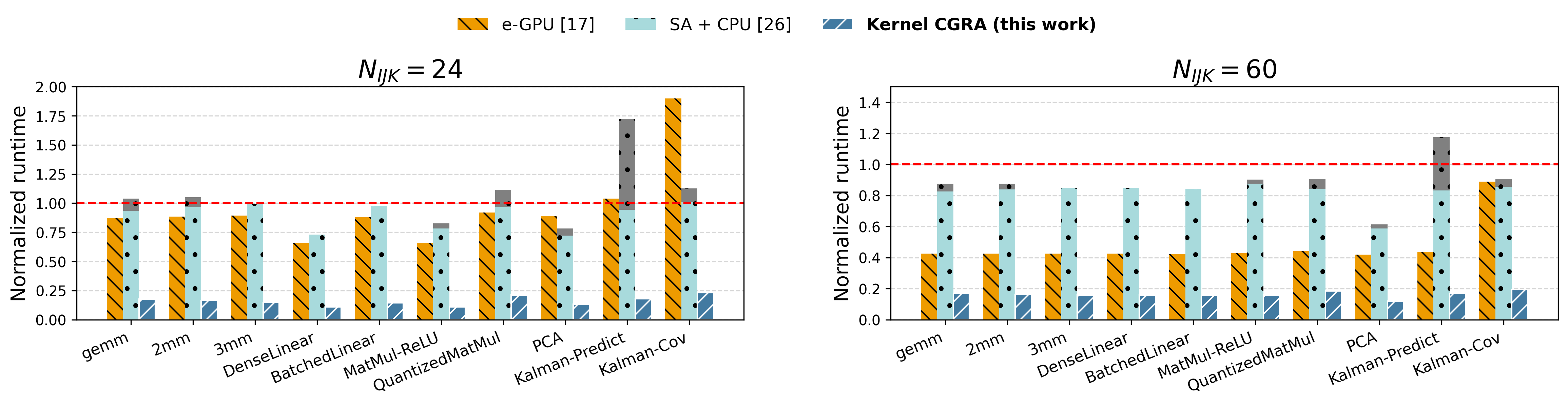}
    \caption{Normalized runtimes on 4x4 OpenEdgeCGRA \cite{openEdgeCGRA} compiled using kernel pre-compilation, e-GPU and systolic array, w.r.t CGRA using CDFG compilation with MS — lower is better.}
    \label{fig:soa comparison}
    \vspace{-1em}
\end{figure}

To evaluate performance against alternative architectural paradigms, we select the $4\times4$ OpenEdgeCGRA configuration, as it represents the area-efficiency sweet spot identified in Section~\ref{subsec:setup}. Figure~\ref{fig:soa comparison} presents the runtime comparison normalized to the baseline CGRA implementation compiled with CDFG mapping. Across all evaluated benchmarks and both data sizes, the proposed \titleacronym-compiled CGRA achieves the lowest runtime. It obtains speedups of $2.5\times$--$7.8\times$ over the e-GPU and $4.8\times$--$7.1\times$ over the SA+CPU configuration, showing that the proposed mapping is competitive not only against compiler-generated CGRA code but also against representative edge accelerators.

The comparison with SA+CPU highlights two different forms of data reuse. The SA improves locality by tiling the output matrix and reusing input data from SRAM across the tile. This organization is effective for large, regular \mmul{} workloads, but its advantage is less pronounced at the edge-scale matrix sizes considered here, where setup, tiling, and CPU/SA orchestration costs are not fully amortized. In contrast, our CGRA kernel exploits the flexibility of the programmable PEs to reuse data directly inside the computation fabric: operands are loaded once and propagated among computing PEs while multiple MAC operations proceed spatially. This reduces pressure on the memory hierarchy and better matches the small-to-medium matrices used in edge applications.

The gap is also affected by programmability outside the pure \mmul{} region. For \textit{DenseLinear}, \textit{MatMul-ReLU}, and \textit{BatchedLinear}, the SA+CPU baseline performs well because most execution time {is spent on} matrix multiplication and the same extracted \mmul{} regions are lowered to the SA. However, for \textit{gemm}, \textit{PCA}, \textit{Kalman-Predict}, and \textit{Kalman-Cov}, non-\mmul{} computations such as scaling, transposition, and covariance updates become significant. SA+CPU executes these residual regions on the host CPU, introducing accelerator-boundary overhead for mixed workloads. Kernel-CGRA instead keeps the residual regions on the CGRA after polyhedral decomposition and schedules them with MS, preserving programmable acceleration outside the dense matrix product. The e-GPU baseline provides a complementary programmable edge platform, but its normalized runtime remains higher because the proposed CGRA kernel is specialized for \mmul{} dataflow while still retaining enough flexibility to handle surrounding non-\mmul{} code.

\vspace{-0.5em}
\section{Ablation study}
\label{sec:ablation}
\subsection{Effect of Affine Normalization on \mmul{} Extraction}

Table~\ref{tab:mmul-ablation} reports the cumulative impact of progressively enabling affine normalization passes on the number of emitted \texttt{cgra.gemm\_blas} operations.
The denominator represents the number of top-level affine loop nests after Polygeist lowering-that is, maximal \texttt{affine.for} nests that are not nested inside another \texttt{affine.for}. 
Among these nests, 13 contain \mmul{}-like computations, while four correspond to non-\mmul{} tasks (scalar, vector, centering, copy, or elementwise operations) and fall outside the scope of \mmul{} extraction.

Scalarization exposes memory-carried reductions, memory canonicalization normalizes layout variants (e.g., \textit{Kalman-Cov} and \textit{PCA}), reduction distribution exposes hidden products (such as \textit{QuantizedMatMul}), and subspace scheduling isolates \mmul{} regions from surrounding scalar work (such as \textit{gemm} and \textit{2mm}). After all passes, all 13 \mmul{}-relevant loop nests are successfully extracted.

\begin{table}[t]
\centering
\caption{{Cumulative ablation of MMUL-region extraction.}}
\vspace{-0.5em}
\label{tab:mmul-ablation}
\begin{tabular}{lc}
\toprule
Configuration & Extracted MMUL Regions / Top-Level Loop Nests \\
\midrule
Pattern matcher & 4/17 \\
+ Scalarization & 8/17 \\
+ Memory canonicalization & 10/17 \\
+ Reduction distribution & 11/17 \\
+ Subspace scheduling/isolation & 13/17 \\
\bottomrule
\end{tabular}
\vspace{-1em}
\end{table}

\subsection{Performance analysis on polyhedral transformation}

\begin{figure}[t]
  \centering
  \includegraphics[width=\columnwidth]{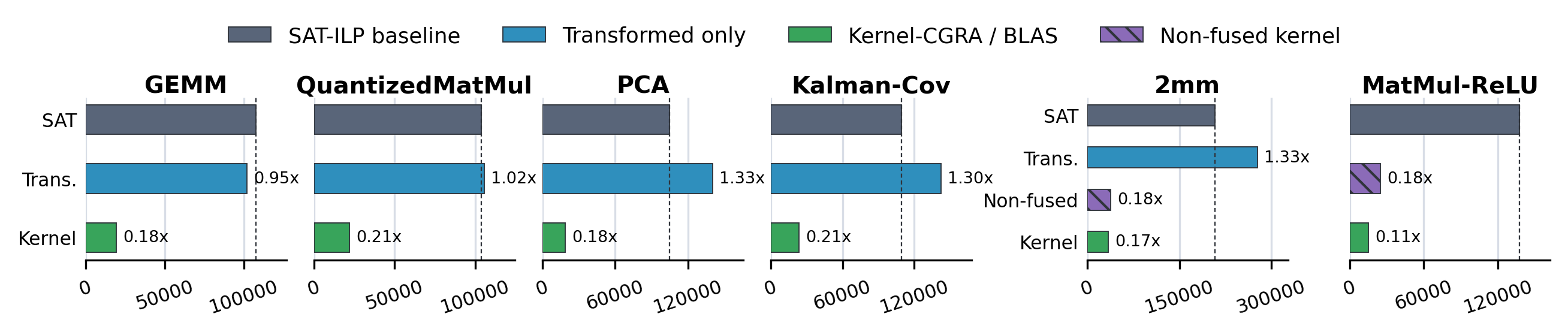}
  \caption{{Effect on runtime of middle-end transformations and back-end assembly substitution.}}
  \vspace{-2em}
  \label{fig:ablation-compiler-split}
\end{figure}

We further perform an ablation study on the selected benchmarks where the \mmul{} pattern cannot be directly extracted by the pattern matcher and therefore requires polyhedral transformation. The goal is to isolate the contribution of middle-end loop transformation from the back-end kernel substitution. For each benchmark, we compare three configurations: baseline SAT-ILP mapping, polyhedral transformation only, and the final Kernel CGRA version with \mmul{} kernel substitution. For \textit{2mm} and \textit{MatMul-ReLU}, we include a non-fused variant to separate the effect of loop fusion from kernel replacement.

Figure~\ref{fig:ablation-compiler-split} shows the cycle count of runtime each configuration. Polyhedral transformation alone does not always guarantee a performance gain. Although it exposes matrix-multiplication-like loop structures, decomposing and restructuring loops may introduce additional control overhead. This effect is visible in \textit{Kalman-Cov} and \textit{2mm}, where the transformed-only version is slower than the SAT-ILP baseline.

However, when the extracted loops reside in higher-level loop nests, the exposed computation can still be modulo scheduled effectively, which compensates for the added control cost. This leads to small improvements in \textit{gemm}. The fused version further improves performance by reducing intermediate control and memory traffic.

\section{Conclusion}
\label{sec:conclusion}

This paper presented Kernel-CGRA, a compilation flow that combines parametric pre-optimized \mmul{} kernels with generic CDFG compilation for CGRA targets. 
The middle-end uses polyhedral transformations to expose legal \mmul{} subspaces within complex loop and access patterns, while the back-end replaces them with optimized kernels and reconnects them to the residual program through context generation and live-value preservation.

The evaluation shows that kernel extraction and integration improve both runtime efficiency and scalability across CGRA sizes and matrix dimensions, indicating that pre-optimized kernels can be integrated into a general CGRA compilation flow without reducing the entire application to a fixed library call interface.

Our implementation is conservative: regions with non-affine accesses, unsupported reductions, ambiguous aliasing, side effects, or unresolved dependences remain in the generic compilation path. The separation between hardware-independent kernel exposure in the middle-end and target-specific kernel generation in the back-end nevertheless provides a structured path for extending the methodology beyond \mmul{}. Future work will explore additional computational kernels through kernel-specific recognition, transformations, and back-end templates, together with architecture-specific optimizations.

\section*{Acknowledgements}
This work was supported in part by the Swiss NSF, grant no. 10.002.812: ”Edge-Companions: Hardware/Software Co-Optimization Toward Energy-Minimal Health Monitoring at the Edge”, and it was partially conducted by ACCESS – AI Chip Center for Emerging Smart Systems, supported by the InnoHK initiative of the Innovation and Technology Commission of the Hong Kong Special Administrative Region Government. This work was partially funded by research grant PID2024-158311NB-I00 by MCIN/AEI/10.13039/501100011033 and by FEDER, EU.

\bibliographystyle{ACM-Reference-Format}
\bibliography{main}


\end{document}